\documentclass[reprint,twocolumn,superscriptaddress,amsmath,amssymb,nofootinbib,preprintnumbers]{revtex4-2} 

\usepackage{mathptmx}
\usepackage{graphicx}  
\usepackage[mathlines]{lineno}
\usepackage[english]{babel} 
\usepackage{xcolor}
\usepackage{ulem}
\usepackage{dcolumn}   
\usepackage{bm}        
\usepackage{amssymb}   
\usepackage{amsbsy}   
\usepackage{comment}
\usepackage{multirow}
\usepackage{diagbox}
\usepackage[draft]{todonotes}
\usepackage{mathtools}
\usepackage{newtxtext}\usepackage[varvw]{newtxmath}
\usepackage{nicefrac}
\usepackage{marginnote}

\usepackage{hyperref}
\hypersetup{
    colorlinks=true,
    linkcolor=cyan,
    filecolor=magenta,      
    urlcolor=cyan,
    citecolor=violet,
}
\usepackage{cleveref}

\makeatletter
\g@addto@macro\bfseries{\boldmath}
\makeatother

\def \dif {\mathrm{d}}
\def \expounit {km$^2$\,sr\,yr}

\begin{document}

\title{The Energy Spectrum of Ultra-High Energy Cosmic Rays across Declinations $-90^\circ$ to $+44.8^\circ$ as measured at the Pierre Auger Observatory}

\author{A.~Abdul Halim}
\affiliation{University of Adelaide, Adelaide, S.A., Australia}

\author{P.~Abreu}
\affiliation{Laborat\'orio de Instrumenta\c{c}\~ao e F\'\i{}sica Experimental de Part\'\i{}culas -- LIP and Instituto Superior T\'ecnico -- IST, Universidade de Lisboa -- UL, Lisboa, Portugal}

\author{M.~Aglietta}
\affiliation{Osservatorio Astrofisico di Torino (INAF), Torino, Italy}
\affiliation{INFN, Sezione di Torino, Torino, Italy}

\author{I.~Allekotte}
\affiliation{Centro At\'omico Bariloche and Instituto Balseiro (CNEA-UNCuyo-CONICET), San Carlos de Bariloche, Argentina}

\author{K.~Almeida Cheminant}
\affiliation{Nationaal Instituut voor Kernfysica en Hoge Energie Fysica (NIKHEF), Science Park, Amsterdam, The Netherlands}
\affiliation{IMAPP, Radboud University Nijmegen, Nijmegen, The Netherlands}

\author{A.~Almela}
\affiliation{Instituto de Tecnolog\'\i{}as en Detecci\'on y Astropart\'\i{}culas (CNEA, CONICET, UNSAM), Buenos Aires, Argentina}
\affiliation{Universidad Tecnol\'ogica Nacional -- Facultad Regional Buenos Aires, Buenos Aires, Argentina}

\author{R.~Aloisio}
\affiliation{Gran Sasso Science Institute, L'Aquila, Italy}
\affiliation{INFN Laboratori Nazionali del Gran Sasso, Assergi (L'Aquila), Italy}

\author{J.~Alvarez-Mu\~niz}
\affiliation{Instituto Galego de F\'\i{}sica de Altas Enerx\'\i{}as (IGFAE), Universidade de Santiago de Compostela, Santiago de Compostela, Spain}

\author{A.~Ambrosone}
\affiliation{Gran Sasso Science Institute, L'Aquila, Italy}

\author{J.~Ammerman Yebra}
\affiliation{Instituto Galego de F\'\i{}sica de Altas Enerx\'\i{}as (IGFAE), Universidade de Santiago de Compostela, Santiago de Compostela, Spain}

% ...

\collaboration{The Pierre Auger Collaboration}
\email{spokespersons@auger.org}
\homepage{http://www.auger.org}
\noaffiliation


\date{\today}

\begin{abstract}
The energy spectrum of cosmic rays above 2.5\,EeV has been measured across the declination range \mbox{$-90^\circ \leq\delta\leq +44.8^\circ$} using $\sim 310{,}000$ events accrued at the Pierre Auger Observatory from an exposure of \mbox{$(104{,}900\pm 3{,}100)$~\expounit}. No significant variations of energy spectra with declination are observed, after allowing or not  for non-uniformities across the sky arising from the well-established dipolar anisotropies in the arrival directions of ultra-high energy cosmic rays. The instep feature in the spectrum at $\simeq$~10\,EeV reported previously is now established at a significance above 5$\,\sigma$. Within the statistics, the energy spectra are indistinguishable across declinations so disfavoring an origin for the instep from a few distinctive sources. 
\end{abstract}

\maketitle

\section{Introduction}
\label{sec:intro}

The discovery at the Pierre Auger Observatory~\cite{PierreAuger:2015eyc} of a dipole in the distribution of cosmic rays of energy above 8\,EeV of amplitude $\sim 6\%$ at a significance of over $5\,\sigma$ marks a significant advance in the field of high-energy cosmic-ray astronomy~\cite{PierreAuger:2017pzq}. The status of this result is now $6.8\,\sigma$, with an increase of the dipole amplitude with energy identified~\cite{PierreAuger:2024fgl}.  Above 32\,EeV, associations of high-energy cosmic rays with an overdensity in the Centaurus region, and with starburst galaxies, have been found, but only at the $4\,\sigma$ level~\cite{PierreAuger:2022axr}. Beyond 100\,EeV, the arrival direction distribution across the whole sky, derived from a combination of data from the Auger Observatory and the Telescope Array, is  featureless~\cite{Fujii:2024sys}. It is thus important to examine whether the energy spectrum varies with declination, particularly at the highest energies, where the new spectral feature near 10\,EeV, identified in data from the Auger Observatory~\cite{PierreAuger:2020qqz}, indicates a two-step suppression of the spectrum~\cite{PierreAuger:2020kuy,Razzaque:2020wdo}. An analysis of spectra in bands of declination is presented, exploiting the 1.2~m deep-water Cherenkov detectors of the Auger Observatory to obtain measurements over declinations $-90^\circ$ to $+44.8^\circ$.

At large zenith angles $\theta$, the effects of the geomagnetic field on the distribution of the muons in air showers increases so that the near-circular symmetry of signals around the shower axis found at low zenith angles is destroyed: above 70$^\circ$, eccentricities over $0.9$ are found. Accordingly, different methods of analyses are necessary for ``vertical showers'' $(\theta\leq 60^\circ)$ and for the more ``inclined events'' ($60^\circ$ to $80^\circ$). The procedures for dealing with events with $\theta\leq 60^\circ$ are classical~\cite{Scherb:1959,Clark:1961mb}. Techniques to reconstruct events where the geomagnetic field is important  were developed more recently~\cite{Ave:2000nd}, with the methods later adapted for use with data from the Auger Observatory. The procedure to obtain the energy of the primary particle for vertical showers is based on the determination of the signal in the water-Cherenkov detectors, $S(1000)$, interpolated at 1000~m from the centre of the shower~\cite{PierreAuger:2020yab}, while for inclined events a parameter based on the number of muons in the shower, $N_{19}$, has been adopted~\cite{Ave:2000xs,PierreAuger:2014jss}. $S(1000)$ and $N_{19}$ are subsequently converted into size parameters $S_{38}$ and $N_{68}$, independent of the zenith angle, using the constant intensity method~\cite{Hersil:1961zz,PierreAuger:2020qqz}. Calorimetric estimates of primary energies are derived by calibrating these size parameters using fluorescence measurements, obtained on clear, moonless nights. Derivation of the energy spectrum from vertical events requires no assumptions about features of hadronic interactions at high energies or of the mass of the primary cosmic rays, apart from a 1.5\% systematic uncertainty in the estimation of the invisible energy~\cite{PierreAuger:2019dhr,PierreAuger:2020qqz}. Minimal assumptions about these parameters are required for the inclined events~\cite{PierreAuger:2015xho,Dembinski:2009jc}. Data used here were recorded between 1 January 2004 and 31 December 2022.  For vertical events the exposure was $\mathcal{E}_{[0-60^\circ]}=(81{,}100\pm2,400)$~\expounit, with $279{,}131$ events recorded above 2.5\,EeV, while for inclined showers $31{,}543$ events above 4\,EeV were recorded in an exposure of $\mathcal{E}_{[60^\circ-80^\circ]}=(23,800\pm 700)$~\expounit. Exposures are independent of energy above these thresholds~\cite{PierreAuger:2020qqz,PierreAuger:2015xho}, which are non-identical mainly due to the attenuation of the electromagnetic cascade at high zenith angles.

\section{Combination of individual spectra}
\label{sec:combination}

\begin{figure}[tp]
\centering
\includegraphics[width=0.5\textwidth]{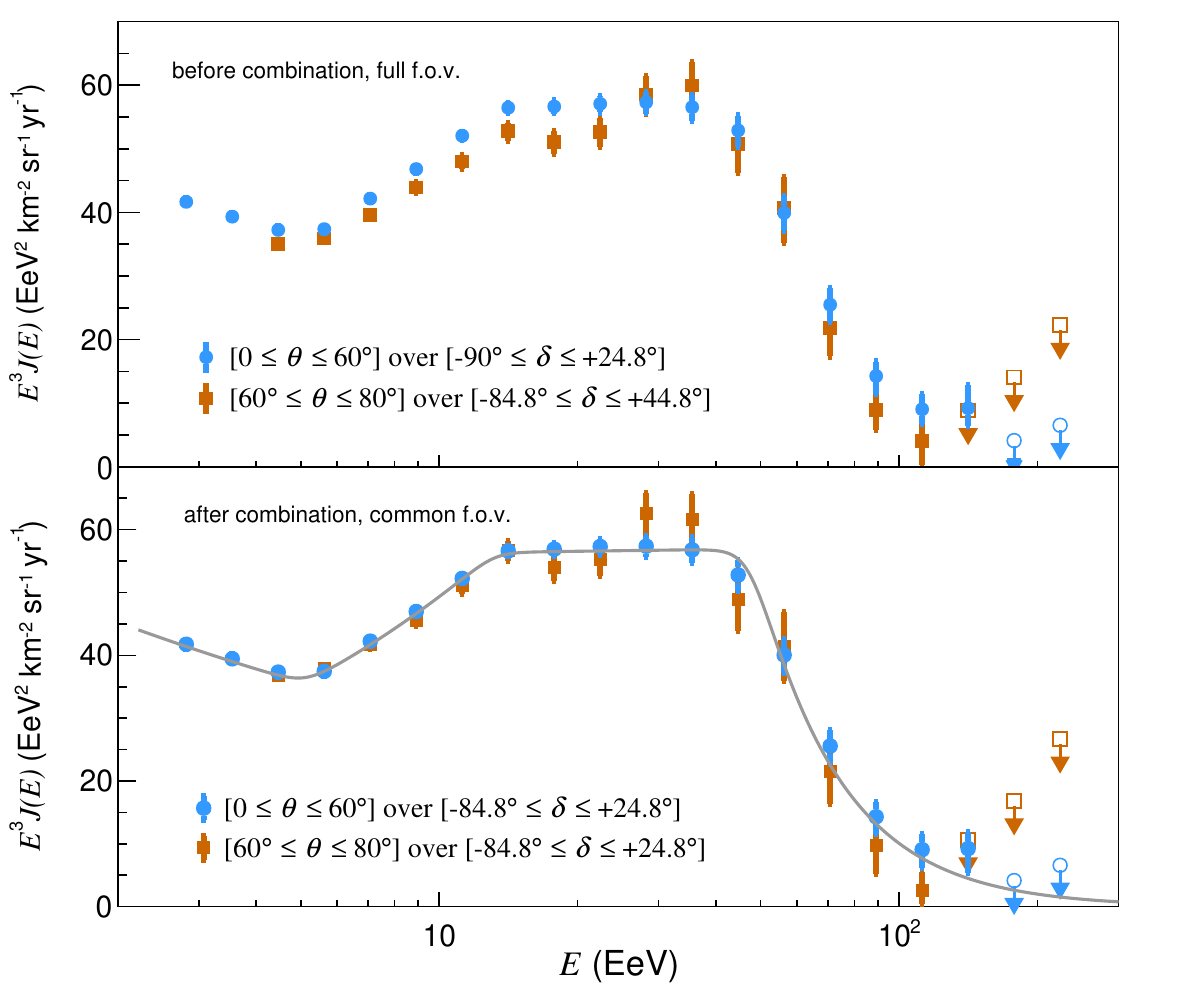}  
\caption{Individual energy spectra scaled by $E^3$ inferred from $S_{38}$- and $N_{68}$-based analyses before (top panel) and after (bottom panel) correcting the energies of inclined events to a best-fit common energy scale. The declination range after combination is reduced to $[-84.8^\circ,+24.8^\circ]$ to guarantee observation of the same sky. Upper limits at 90\% confidence level are shown in empty bins. The gray line is the best-fit to data points after combination.}
\label{fig:spectra_commonband}
\end{figure}

Given the latitude of the Observatory, $\lambda\simeq -35.2^\circ$ and the zenith angle ranges mentioned in the previous section, the $S_{38}$-based dataset covers the range of declinations $-90^\circ\leq \delta \leq +24.8^\circ$ in equatorial coordinates, whereas the $N_{68}$ one covers $-84.8^\circ \leq \delta \leq +44.8^\circ$. Our aim is to measure the spectrum over the whole declination range covered with the surface array, $-90^\circ \leq \delta \leq +44.8^\circ$. Thus the independent spectra from the $S_{38}$- and $N_{68}$-based analyses, shown as the circles and squares respectively in the top panel of Fig.~\ref{fig:spectra_commonband}, must be combined.

The combination procedure follows from that in~\cite{PierreAuger:2021hun}. The combined spectrum is obtained by forward-folding the effects of the finite energy accuracy of the detector into a proposed function tailored to describe a series of power-law falloffs,
\begin{equation}
\label{eqn:J}
J(E;\mathbf{p}) = J_0 \left(\frac{E}{E_0}\right)^{-\gamma_0} \frac{\prod_{j=1}^3\left[1+\left(\frac{E}{E_{jk}}\right)^{\omega_{jk}^{-1}}\right]^{(\gamma_j-\gamma_{k})\omega_{jk}}}{\prod_{j=1}^3\left[1+\left(\frac{E_0}{E_{jk}}\right)^{\omega_{jk}^{-1}}\right]^{(\gamma_j-\gamma_{k})\omega_{jk}}},
\end{equation}
with $k=j+1$ and eight free parameters encompassed in $\mathbf{p}$, namely the overall normalization $J_0$, four spectral indices $(\gamma_0,\gamma_1,\gamma_2,\gamma_3)$ and three energy turning points $E_{jk}$; the parameters $\omega_{jk}$ that govern the width of the transition from $\gamma_j$ to $\gamma_{k}$ are fixed to $\omega_{jk}=0.05$~\cite{PierreAuger:2020qqz}; the pivot energy is chosen as $E_0=10^{0.5}~$EeV. The parameters $\mathbf{p}$ are adjusted to obtain the best match between the observed number of events $n_i$ in each differential energy bin of width $\Delta\log_{10}E=0.1$ and the expected ones $\nu_i(\mathbf{p})$ simultaneously for both data streams. The minimization procedure is based on the product of Poissonian likelihood functions pertaining to each individual spectrum, $\mathcal{L}_{[0-60^\circ]}$ and $\mathcal{L}_{[60^\circ-80^\circ]}$. 
  
Both individual spectra are subject to systematic uncertainties, with contributions from the absolute energy scale ($14\%$)~\cite{Verzi:2013ajy}, the exposure ($3\%$)~\cite{PierreAuger:2010zof}, the unfolding procedure ($\leq 2\%$)~\cite{PierreAuger:2020qqz} and the energies inferred from $S_{38}$ and $N_{68}$ ($\leq 3\%$)~\cite{PierreAuger:2020qqz,PierreAuger:2014jss}. No indication of further systematics has been found from a comparison of spectra calculated over different time periods, seasons and ranges of zenith angle. 

The systematic uncertainties relating to each $S_{38}$- and $N_{68}$-based analysis are common to both: the important exception concerns those inherited from the energy-calibration procedure that are uncorrelated as the two datasets used for calibration are independent. The largest uncertainties are related to the calibration of $N_{68}$, as it is based on $605$ events compared to $4{,}703$ for $S_{38}$. The statistics is limited by the high-quality criteria used to guarantee an accurate energy estimation with the fluorescence technique, and by the effective area for observing the maximum of shower development that drops rapidly with zenith angle~\cite{Guerard:2002yg,PierreAuger:2010swb}. Therefore, our strategy for the combination is to correct the energies of the $N_{68}$-based dataset, originally determined as $E=A{N_{68}}^B$, to a best-fit common energy scale, using two absolute departure parameters $\delta A$ and $\delta B$ such that $E'=(A+\delta A){N_{68}}^{B+\delta B}$. As shown below, considering the uncertainties in $N_{68}$ alone, the most significant ones, will suffice to obtain a satisfactory goodness-of-fit. Including those in $S_{38}$ would improve further the combination, but is unnecessary. 

In addition, as explained in the End Matter, the energy calibration of $N_{68}$ may be affected by (logarithmic) non-linearities above $\sim 10~$EeV due to the sensitivity of $N_{68}$ to the small, yet sharp, changes in mass composition recently uncovered~\cite{PierreAuger:2024flk,PierreAuger:2024nzw} and to experimental effects. A third departure parameter, $\delta C$, is therefore introduced so that the correction reads $E''=(A+\delta A){N_{68}}^{B+\delta B+\delta C}$ above $10~$EeV.  

During each step of the fit, the energy-bin boundaries of the $N_{68}$-based analysis are corrected by varying the extra-parameters $\mathbf{x}=(\delta A,\delta B,\delta C)$ that govern the uncorrelated systematic uncertainties. Correspondingly, an extra-term in the joint likelihood restricts the values of the extra-parameters within their uncertainties. It is constructed by considering on the one hand the sum of two random variables $\delta B$ and $\delta C$, and on the other hand the correlation between $\delta A$ and $\delta B$. The log-likelihood function therefore stems from the convolution of a 2D Gaussian with correlation parameter $\rho\simeq -0.66$ with a 1D Gaussian pertaining to $\delta C$ alone,
\begin{multline}
    -2\ln{\mathcal{L}_\mathbf{x}(\delta A,\delta B,\delta C})=\\
    \frac{\delta A^2(\sigma_{B}^2+\sigma_{C}^2)-2\rho\delta A\delta B\sigma_{A}\sigma_{B}+(\delta B+\delta C)^2\sigma_{A}^2}{\sigma_{A}^2(\sigma_{C}^2+\sigma_{B}^2(1-\rho^2))},
\end{multline}
with $\sigma_{A}\simeq6\times 10^{-2}~$EeV, $\sigma_{B}= 1.4\times10^{-2}$ and, as shown in the End Matter, $\sigma_C=3\times10^{-2}$ from the systematic uncertainties in $N_{68}$ and in the relationship between $N_{68}$ and energy. 

In this way, for unrealistic changes in $(\delta A,\delta B,\delta C)$, the $-2\ln{\mathcal{L}_\mathbf{x}(\delta A,\delta B,\delta C)}$ term acts as a penalty factor while minimizing the total log-likelihood function $-2\ln{\mathcal{L}(\mathbf{p},\mathbf{x})}$, which, overall, reads as
\begin{equation}
    \label{eqn:L}
    \mathcal{L}(\mathbf{p},\mathbf{x})=\mathcal{L}_{[0-60^\circ]} (\mathbf{p},\mathbf{x})\times\mathcal{L}_{[60^\circ-80^\circ]}(\mathbf{p},\mathbf{x})\times \mathcal{L}_\mathbf{x}(\mathbf{x}).
\end{equation}

The outcome of the forward-folding is the set of parameters $\mathbf{p}$, and $\mathbf{x}$, that allows calculation of $\nu_i$ and $\mu_i$, which are the expected number of events including and not including the detector effects, respectively. Unfolding factors, defined as $c_i=\mu_i/\nu_i$, are then applied to correct for bin-to-bin migration induced by the finite accuracy of the response functions, which are determined in a data-driven manner~\cite{PierreAuger:2020qqz} and given in the Supplemental material. The resulting spectral point for each bin is obtained as
\begin{equation}
    \label{eqn:Jcomb}
    J_i=\frac{c_i n_{i[0-60^\circ]}+c_in_{i[60^\circ-80^\circ]}}{\mathcal{E}_i\Delta E_i},
\end{equation}
with $\mathcal{E}_i=\mathcal{E}_{[0-60^\circ]}$ for $0.4\leq \log_{10}{(E/\mathrm{EeV})}\leq 0.6$ and $\mathcal{E}_i=\mathcal{E}_{[0-60^\circ]}+\mathcal{E}_{[60^\circ-80^\circ]}$ for $\log_{10}{(E/\mathrm{EeV})}\geq 0.6$. The minimization procedure, applied only to the events in the declination band $[-84.8^\circ,+24.8^\circ]$, common to both datasets, to guarantee observation of the same sky so that spectra must be in statistical agreement, yields values of $\delta A/A\simeq (3.0\pm0.6)\%$, $\delta B/B\simeq (0.3\pm1.5)\%$, respectively corresponding to shifts of $2.6~\sigma_A$ and $0.2~\sigma_B$, and $\delta C\simeq (-2.0\pm 2.0)\times 10^{-2}$. The resulting spectra are displayed in the lower panel of Fig.~\ref{fig:spectra_commonband}. The most significant changes are in the lower energy bins, where the statistical power is greatest. Note that, as the analysis is designed so that $\mathcal{L}_{[0-60^\circ]} (\mathbf{p},\mathbf{x})$ is independent of $\mathbf{x}$, the changes impact dominantly the $N_{68}$-based spectrum. The combined spectrum obtained with eqn.~\ref{eqn:Jcomb} shows a deviance of $D=40.5$, which, if considered to follow a ``C statistics''~\cite{Bonamente:2019efn}, can be compared to the expectation of $\langle D\rangle=26.9\pm7.0$ to yield a p-value of $\simeq 0.12$.

\section{Searches for declination dependences}
\label{sec:dec_dependencies}

\begin{table*}[t]
\caption{Spectral parameters $\mathbf{p}$ in several declination bands: normalization $J_0$ in km$^{-2}$\,sr$^{-1}$\,yr$^{-1}$\,EeV$^{-1}$ units, 
indices $\gamma_j$ and break energies $E_{jk}$ in EeV units; and probabilities $P(\geq Q^2)$ that $\mathbf{p}_{\Delta\delta}$ departs from $\mathbf{p}_\mathrm{ref}$ measured across $[-84.8^\circ,+24.8^\circ]$. In the declination band $[+24.8^\circ,+44.8^\circ]$, the threshold of the $N_{68}$-based analysis does not allow for measuring $\gamma_1$ and $E_{12}$, which are fixed to their value found across $[-84.8^\circ,+24.8^\circ]$.
}
\label{tab:features}
\begin{ruledtabular}
\begin{tabular}{l c c c c c c c c c} 
$[\delta_\mathrm{min},\delta_\mathrm{max}]$ & $J_0$ & $\gamma_1$ & $\gamma_2$ & $\gamma_3$ & $\gamma_4$ & $E_{12}$ & $E_{23}$ & $E_{34}$ & $P(\geq Q^2)$ \\ 
\colrule
&\\[-1.0em]
$[-84.8^\circ,+24.8^\circ]$ & $1.271\pm0.004$ & $3.26 \pm 0.01$ & $2.51 \pm 0.03$ & $2.99 \pm 0.03$ & $5.3 \pm 0.2$ & $5.1 \pm 0.1$ & $13 \pm 1$ & $48 \pm 2$ & $-$ \\ 
$[-90^\circ,-51^\circ]$ & $1.278\pm0.007$ & $3.24 \pm 0.02$ & $2.54 \pm 0.06$ & $3.18 \pm 0.06$ & $ 7.2\pm 1.0$ & $5.1 \pm 0.2$ & $ 17 \pm 2$ & $ 62 \pm 4$ & 23\% \\ 
$[-51^\circ,-29^\circ]$ & $1.281\pm0.007$ & $3.26 \pm 0.02$ & $2.46 \pm 0.06$ & $2.87 \pm 0.06$ & $4.6 \pm 0.4$ & $5.2 \pm 0.2$ & $12 \pm 2$ & $39 \pm 4$ & 13\% \\ 
$[-29^\circ,-8^\circ]$ & $1.257\pm0.007$ & $3.28 \pm 0.02$ & $2.54 \pm 0.06$ & $3.02 \pm 0.06$ & $5.6 \pm 0.4$ & $4.9 \pm 0.2$ & $13 \pm 2$ & $49 \pm 4$ & 71\% \\ 
$[-8^\circ,+24.8^\circ]$ & $1.266\pm0.007$ & $3.26 \pm 0.02$ & $2.50 \pm 0.06$ & $2.97 \pm 0.06$ & $ 6.1 \pm 0.4$ & $5.1 \pm 0.2$ & $12 \pm 2$ & $51 \pm 4$ & 34\% \\ 
$[+24.8^\circ,+44.8^\circ]$ & $1.26\pm0.03$ & $3.26$ (fixed) & $2.6 \pm 0.2$ & $3.0 \pm 0.2$ & $ 11 \pm 10$ & $5.1$ (fixed) & $13 \pm 4$ & $69 \pm 20$ & 53\% \\ 
\end{tabular}
\end{ruledtabular}
\end{table*}

\begin{figure}[tp]
\centering
\includegraphics[width=0.5\textwidth]{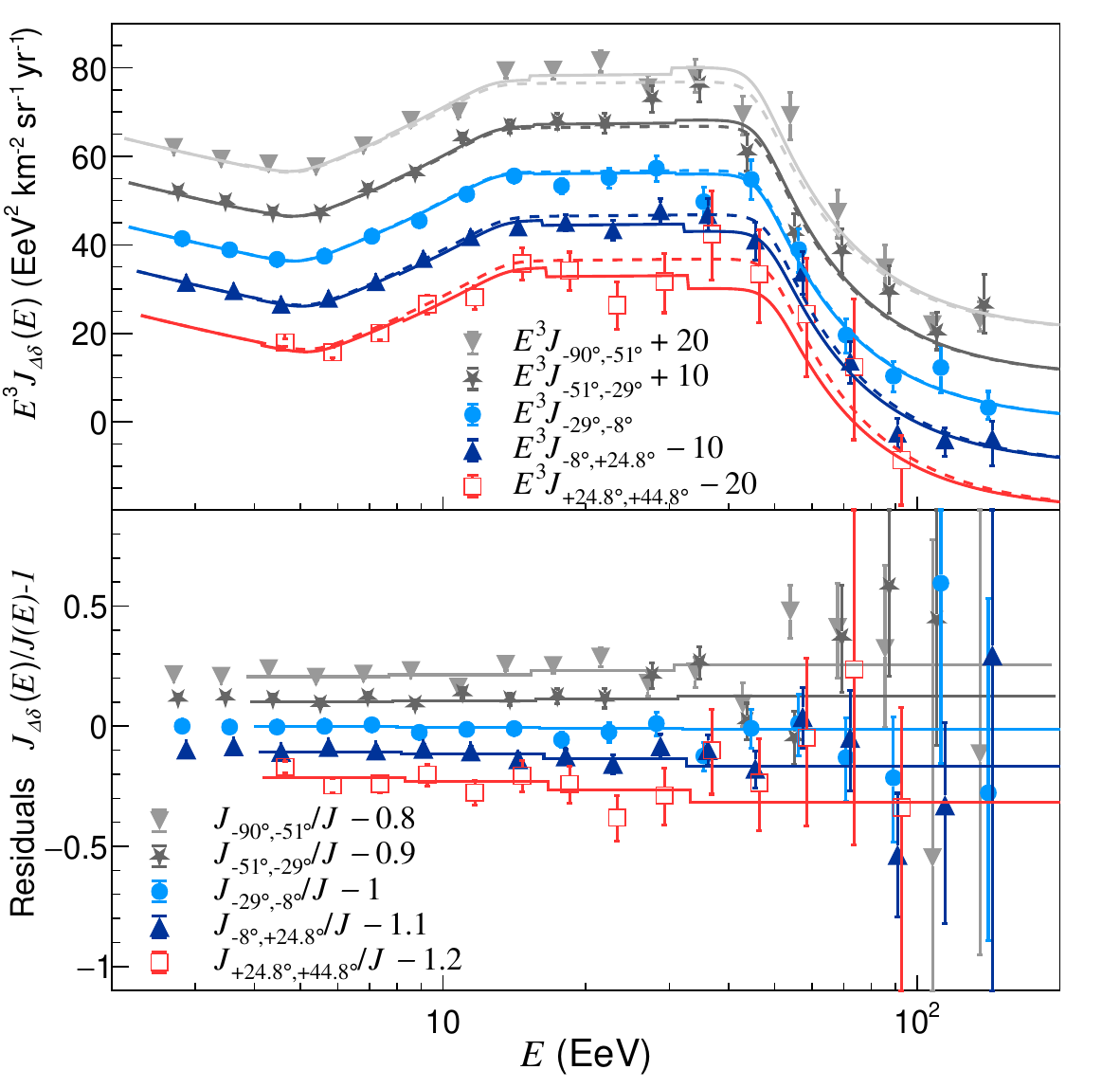}  
\caption{Top: Energy spectra in five declination ranges. The dotted reference lines are the best-fit function for the spectrum combined over $[-84.8^\circ,+24.8^\circ]$; the full lines account for the impact of dipole anisotropies in each band. Bottom: Corresponding residuals. Artificial shifts are applied for visualization purpose. An alternative view of the residuals is provided in the Supplemental material.}
\label{fig:residuals}
\end{figure}

The best-fit parameters for $(\delta A, \delta B,\delta C)$ that yield statistical agreement of spectra in the declination range $[-84.8^\circ,+24.8^\circ]$ are used to search for declination dependences in five ranges over $[-90^\circ,+44.8^\circ]$. The first range corresponds to the most northerly declination band $[+24.8^\circ, +44.8^\circ]$: it is selected as it only contains inclined events ($5{,}632$ after energy corrections for an exposure of $(4{,}100\pm 120)$~\expounit) and it is thus not considered in the combination fit. The remaining portion of the sky $[-90^\circ,+24.8^\circ]$ is divided in four declination ranges with exposures similar to within 1\% and averaging $25{,}000$~\expounit. Detailed comparisons of the spectra in the five declination bands are shown in Fig.~\ref{fig:residuals}. In the top panel, individual spectra are compared with the best-fit function (dotted lines) for the spectrum combined over $[-84.8^\circ,+24.8^\circ]$ and with the same best-fit function (full lines) taking into account the variation of the dipole amplitude with energy and declination measured above 4\,EeV. Because the energy-spectrum estimator is based primarily on the observed number of events through the directional exposure function $\omega(\delta)$, the expression of $J_{\Delta\delta}(E;\mathbf{p})$ is obtained as~\cite{PierreAuger:2020qqz},
\begin{multline}
J_{\Delta\delta}(E;\mathbf{p})=J(E;\mathbf{p})\bigg(1+\\
\frac{\mathcal{E}_0}{\mathcal{E}_{\Delta\delta}}\frac{\int_{\Delta\delta}\dif\delta\cos{\delta}~\omega(\delta)+d_z(E)\int_{\Delta\delta}\dif\delta\cos{\delta}\sin{\delta}~\omega(\delta)}{\int\dif\delta\cos{\delta}~\omega(\delta)+d_z(E)\int\dif\delta\cos{\delta}\sin{\delta}~\omega(\delta)}\bigg),
\end{multline}
with $\mathcal{E}_0$ the total exposure over the declination range of reference $[-84.8^\circ,+24.8^\circ]$,  $\mathcal{E}_{\Delta\delta}$ that over the declination band under consideration, and where the integrations in the denominator are carried out over $[-84.8^\circ,+24.8^\circ]$. The coefficients $d_z(E)$ read $\{-0.013,-0.031,-0.070\}$ in differential bins of width $\Delta\log_{10}E=0.3$ between 4 and 32\,EeV and $-0.13$ above 32\,EeV~\cite{PierreAuger:2024fgl}. 
In the bottom panel, the best-fit function over $[-84.8^\circ,+24.8^\circ]$ is taken as a reference and the residual differences from this spectrum are compared with expectations. The residuals follow the trend imprinted by the dipole between 4 and 32\,EeV; statistical fluctuations dominate above 32 EeV. 

Spectral parameters for each declination band, $\mathbf{p}_{\Delta\delta}$, are obtained by applying the combination procedure, except in the northernmost band where only inclined events contribute: they are listed in Table~\ref{tab:features}. A quantitative statement on the statistical agreement between $\mathbf{p}_{\Delta\delta}$ and the reference parameters across $[-84.8^\circ,+24.8^\circ]$, $\mathbf{p}_\mathrm{ref}$, can be drawn from the distance-squared quantity
\begin{equation}
    \label{eqn:Q2}
    Q^2=(\mathbf{p}_{\Delta\delta}-\mathbf{p}_\mathrm{ref})^\mathrm{T}(\mathbf{\Sigma}_{\Delta\delta}+\mathbf{\Sigma}_{\mathrm{ref}})^{-1}(\mathbf{p}_{\Delta\delta}-\mathbf{p}_\mathrm{ref}),
\end{equation}
which accounts for the correlations between parameters through their respective covariance matrices $\mathbf{\Sigma}_{\Delta\delta}$ and $\mathbf{\Sigma}_{\mathrm{ref}}$ given in the Supplemental material. Some fraction of the uncertainty in the exposure may depend on the declination~\cite{PierreAuger:2012xju}. The array is slightly tilted by $\simeq 0.2^\circ$ in the southeastern direction $\varphi_\mathrm{tilt} \simeq -30^\circ$. Besides, the latitude of the Observatory varies by $\simeq 0.5^\circ$ between detectors located in the northernmost and southernmost positions. Considering the uncertainties in the determination of $\varphi_\mathrm{tilt}$ and the impact of the spatial extension of the array, we conservatively allocate an uncertainty of 100~\expounit~  in the exposure of each declination band that propagates into an additional contribution to the uncertainty in $J_0$. The probability that the set of parameters $\mathbf{p}_{\Delta\delta}$ departs from the reference ones is obtained by drawing at random mock samples and by counting the fraction of them displaying a value $Q^2$ larger than that in data. No significant departure is observed.
 
Some departure in $\gamma_3$, $E_{34}$ and $\gamma_4$, which shape the spectrum at the highest energies, are evident in the band $[-51^\circ,-29^\circ]$. The departure remains quasi-identical when fixing the other parameters to their values in the band $[-84.8^\circ, +24.8^\circ]$. This declination range encompasses the Centaurus region in which an overdensity over a circular region of $\simeq 24^\circ$ is present~\cite{PierreAuger:2022axr} and which has attracted attention in, e.g., ~\cite{Biermann:2011wf,Keivani:2014kua,Bell:2021pkk,Mollerach:2024sjd,Marafico:2024qgh,Mbarek:2024nvv}. We have estimated the spectral features by restricting further, within the declination band, the region of interest within $\simeq 24^\circ$ around the position $(\mathrm{right~ascension~}\alpha=201^\circ,~\mathrm{declination~}\delta=-43^\circ)$ to check whether the harder values of $\gamma_3$ and $\gamma_4$ as well as the smaller value of $E_{34}$ elsewhere can be attributed to the overdensity. However, statistical fluctuations prevent drawing any conclusion, given that $\gamma_3=2.8 \pm 0.3$, $E_{34}=(67\pm11)~$EeV and $\gamma_4=9\pm4$ in that specific region. In parallel, the values obtained by excluding the $24^\circ$ window around the Centaurus region from the declination band are $\gamma_3=2.88\pm0.07$, $E_{34}=(37\pm4)$\,EeV and $\gamma_4=4.5\pm0.5$, close to those observed over the whole $[-51^\circ:-29^\circ]$ band.  The same outcome holds about other, more modest, overdensities over the same scale that have attracted attention~\cite{Romero:2018mnb,Anchordoqui:2019mfb,Bell:2021pkk,Marafico:2024qgh}: $\gamma_3=2.9 \pm 0.2$, $E_{34}=(41\pm11)~$EeV and $\gamma_4=5\pm1$ around the region of NGC~253 $(\alpha=11.9^\circ,\delta=-25.3^\circ)$, and, within the declination band $[-8^\circ,+24.8^\circ]$, $\gamma_3=3.3 \pm 0.3$, $E_{34}=(58\pm18)~$EeV and $\gamma_4=5\pm1$ around the region of NGC~1068 galaxy $(\alpha=40.5^\circ,\delta=0^\circ)$. Finally, since the northernmost declination band is also observed using Telescope Array with an exposure $\simeq 20\%$ larger~\cite{TelescopeArray:2024tbi}, it is interesting to note that the spectral parameters do not show deviations with respect to those of the $[-84.8^\circ, +24.8^\circ]$ band. 

We conclude, consistent with large-scale anisotropy measurements~\cite{PierreAuger:2024fgl} and with our previous report from an exposure of $60{,}400~$\expounit~ and covering declinations from $-90^\circ$ to $+24.8^\circ$~\cite{PierreAuger:2020kuy}, that the energy spectra are identical to within a mild dipolar modulation with declination from $-90^\circ$ to $+44.8^\circ$.  

\section{Spectrum across declinations $-90^\circ$ to $+44.8^\circ$}
\label{sec:three_quarter}

\begin{figure}[tp]
\centering
\includegraphics[width=0.5\textwidth]{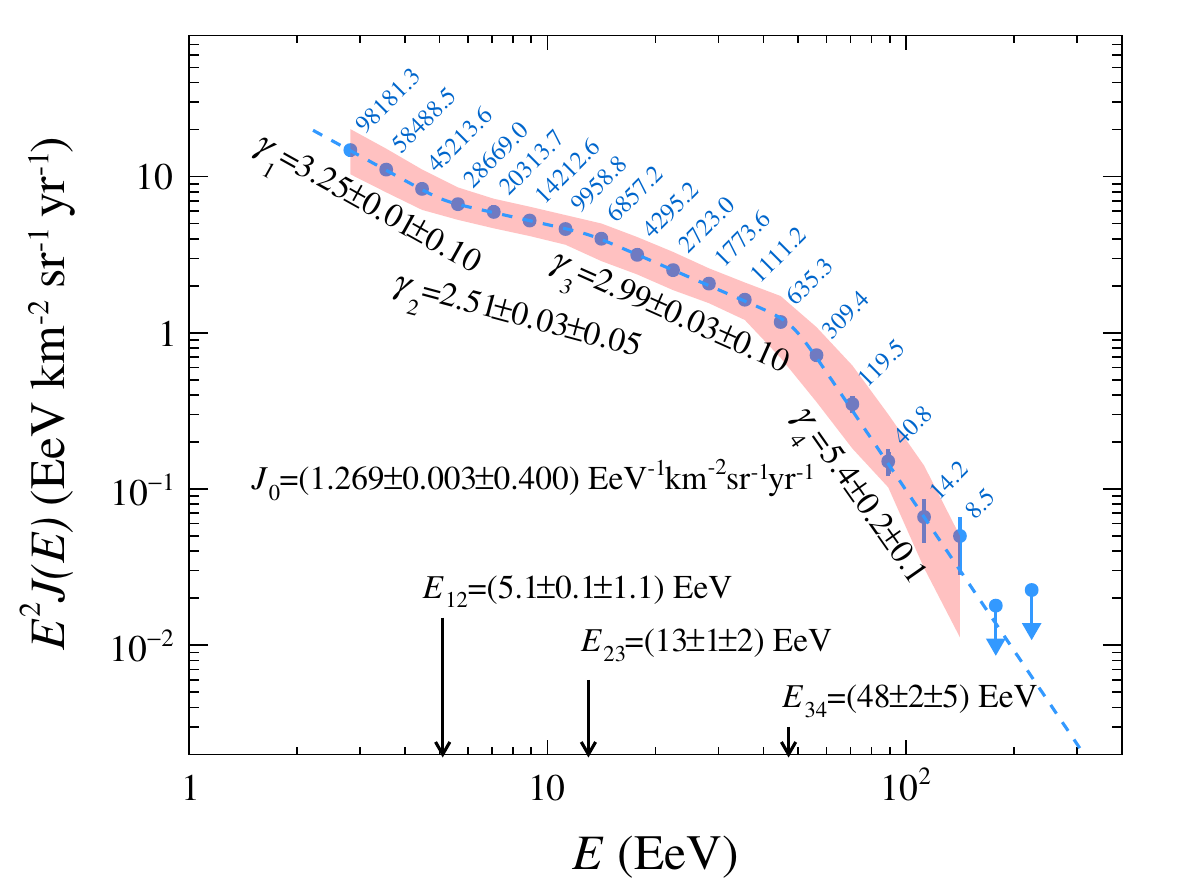}  
\caption{Energy spectrum scaled by $E^2$ (energy flux) across declinations $[-90^\circ,+44.8^\circ]$. The number of events corrected for detector effects is indicated in each bin. The red band stands for the systematic uncertainties while the dotted line indicates the best-fit function described by the spectral features given with their statistical and systematic uncertainties. }
\label{fig:spectrum}
\end{figure}

Given the statistical agreement of the spectra across declinations, we present in Fig.~\ref{fig:spectrum} the combined spectrum across declinations $-90^\circ$ to $+44.8^\circ$, scaled by $E^2$ (energy flux). The colored band shows the systematic uncertainties, which have been propagated by repeating the analysis on datasets obtained by randomly sampling the different sources of systematics that affect individual energies and exposure. The main contribution stems, by far, from the $14\%$ uncertainty in energy scale. The various spectral features, with their statistical and systematic uncertainties, are given in Fig.~\ref{fig:spectrum}. The data, included those in each declination band, are listed on a point-by-point basis in the Supplemental material. 

Beyond the well-known features of the ankle and the steepening, earlier evidence for ``instep''~\cite{PierreAuger:2020qqz} is reinforced with this measurement. We had previously disfavored, with $3.9~\sigma$ confidence, a model of the energy spectrum consisting of a series of two power laws followed by a slow suppression~\cite{PierreAuger:2020qqz}. By drawing at random mock samples of observed energies following this reference model without instep, and by reconstructing the corresponding spectra using the reference model and the alternative one (eqn.~\ref{eqn:J}), we built the distribution of the test statistics~$t$ from the ratio of the likelihood values associated with each hypothesis. Only two out of $10^8$ realizations show a test statistic greater than the observed one ($t\simeq38$). Thus for 1-sided Gaussian distributions, the reference model can be rejected at the $5.5~\sigma$ confidence level. The significance remains above 5.5$~\sigma$ when the complete chain of analysis is repeated with the energy scale adjusted by $\pm14\%$ (the dominant systematic uncertainty) from the nominal one.

\section{Discussion}
\label{sec:discussion}

The first measurement of the energy spectrum of cosmic rays over the entire range of declination covered by the surface array of the Pierre Auger Observatory with an exposure of $104{,}900$~\expounit\, has been presented. This is made possible by using independent spectra  measured over $[0,60^\circ]$ and $[60^\circ,80^\circ]$ combined through a forward-folding procedure that makes use of a unique function to describe the underlying spectrum. Statistical agreement is obtained by increasing the energies inferred from $N_{68}$ values measured between $60^\circ$ and $80^\circ$ from $+2.9$\% at 4\,EeV to $3.1$\% at 10\,EeV; the increase then slows and energies are changed by $-1\%$ at 100\,EeV. Such changes are consistent with the uncertainties in the energy assignments to $N_{68}$. 

A search for a declination dependence was performed by comparing the energy spectrum in five declination ranges with that measured in the band $[-84.8^\circ,+24.8^\circ]$, which is covered when measuring individual spectra between $[0,60^\circ]$ and $[60^\circ,80^\circ]$ in zenith angle. All spectra are found to be consistent with the reference one, apart from the mild modulation expected from a dipolar anisotropy previously uncovered~\cite{PierreAuger:2024fgl}. Notably this statement applies to the northernmost declination band $[+24.8^\circ,+44.8^\circ]$, where only inclined events are available. 

The combination of the two individual measurements and the statistical agreement across declinations lead to the construction of the spectrum across $-90^\circ$ to $+44.8^\circ$ declinations, where the features of the ankle, the instep and the suppression are firmly established. Among those features, the significance of the instep, originally uncovered with $3.9~\sigma$ confidence~\cite{PierreAuger:2020qqz}, has reached 5.5~$\sigma$. The increase in significance is consistent with the increase in exposure, in particular from the inclined events.  

The absence of declination dependencies disfavors that the origin of the instep feature may be attributed to the distinctive spectrum of a small number of foreground sources contributing significantly to the total intensity. By contrast, the steepening seems to reflect the interplay between the flux contributions of the helium and carbon-nitrogen-oxygen components from rather similar sources~\cite{PierreAuger:2020kuy,PierreAuger:2022atd}. This is inline with the narrow range of maximum rigidity at the sources that can be inferred from the succession of rather pure-composition components of nuclei above 10\,EeV~\cite{Ehlert:2022jmy,Farrar:2024zsm,Zhang:2024sjp}. The improved sensitivity in mass-composition with the upgraded Observatory~\cite{PierreAuger:2016qzd} will  allow further characterization on the instep thus shedding more light on its origin.

\textit{Acknowledgments.} The successful installation, commissioning, and operation of the Pierre
Auger Observatory would not have been possible without the strong
commitment and effort from the technical and administrative staff in
Malarg\"ue. We are very grateful to the following agencies and
organizations for financial support:

\begin{sloppypar}
Argentina -- Comisi\'on Nacional de Energ\'\i{}a At\'omica; Agencia Nacional de
Promoci\'on Cient\'\i{}fica y Tecnol\'ogica (ANPCyT); Consejo Nacional de
Investigaciones Cient\'\i{}ficas y T\'ecnicas (CONICET); Gobierno de la
Provincia de Mendoza; Municipalidad de Malarg\"ue; NDM Holdings and Valle
Las Le\~nas; in gratitude for their continuing cooperation over land
access; Australia -- the Australian Research Council; Belgium -- Fonds
de la Recherche Scientifique (FNRS); Research Foundation Flanders (FWO),
Marie Curie Action of the European Union Grant No.~101107047; Brazil --
Conselho Nacional de Desenvolvimento Cient\'\i{}fico e Tecnol\'ogico (CNPq);
Financiadora de Estudos e Projetos (FINEP); Funda\c{c}\~ao de Amparo \`a
Pesquisa do Estado de Rio de Janeiro (FAPERJ); S\~ao Paulo Research
Foundation (FAPESP) Grants No.~2019/10151-2, No.~2010/07359-6 and
No.~1999/05404-3; Minist\'erio da Ci\^encia, Tecnologia, Inova\c{c}\~oes e
Comunica\c{c}\~oes (MCTIC); Czech Republic -- GACR 24-13049S, CAS LQ100102401,
MEYS LM2023032, CZ.02.1.01/0.0/0.0/16{\textunderscore}013/0001402,
CZ.02.1.01/0.0/0.0/18{\textunderscore}046/0016010 and
CZ.02.1.01/0.0/0.0/17{\textunderscore}049/0008422 and CZ.02.01.01/00/22{\textunderscore}008/0004632;
France -- Centre de Calcul IN2P3/CNRS; Centre National de la Recherche
Scientifique (CNRS); Conseil R\'egional Ile-de-France; D\'epartement
Physique Nucl\'eaire et Corpusculaire (PNC-IN2P3/CNRS); D\'epartement
Sciences de l'Univers (SDU-INSU/CNRS); Institut Lagrange de Paris (ILP)
Grant No.~LABEX ANR-10-LABX-63 within the Investissements d'Avenir
Programme Grant No.~ANR-11-IDEX-0004-02; Germany -- Bundesministerium
f\"ur Bildung und Forschung (BMBF); Deutsche Forschungsgemeinschaft (DFG);
Finanzministerium Baden-W\"urttemberg; Helmholtz Alliance for
Astroparticle Physics (HAP); Helmholtz-Gemeinschaft Deutscher
Forschungszentren (HGF); Ministerium f\"ur Kultur und Wissenschaft des
Landes Nordrhein-Westfalen; Ministerium f\"ur Wissenschaft, Forschung und
Kunst des Landes Baden-W\"urttemberg; Italy -- Istituto Nazionale di
Fisica Nucleare (INFN); Istituto Nazionale di Astrofisica (INAF);
Ministero dell'Universit\`a e della Ricerca (MUR); CETEMPS Center of
Excellence; Ministero degli Affari Esteri (MAE), ICSC Centro Nazionale
di Ricerca in High Performance Computing, Big Data and Quantum
Computing, funded by European Union NextGenerationEU, reference code
CN{\textunderscore}00000013; M\'exico -- Consejo Nacional de Ciencia y Tecnolog\'\i{}a
(CONACYT) No.~167733; Universidad Nacional Aut\'onoma de M\'exico (UNAM);
PAPIIT DGAPA-UNAM; The Netherlands -- Ministry of Education, Culture and
Science; Netherlands Organisation for Scientific Research (NWO); Dutch
national e-infrastructure with the support of SURF Cooperative; Poland
-- Ministry of Education and Science, grants No.~DIR/WK/2018/11 and
2022/WK/12; National Science Centre, grants No.~2016/22/M/ST9/00198,
2016/23/B/ST9/01635, 2020/39/B/ST9/01398, and 2022/45/B/ST9/02163;
Portugal -- Portuguese national funds and FEDER funds within Programa
Operacional Factores de Competitividade through Funda\c{c}\~ao para a Ci\^encia
e a Tecnologia (COMPETE); Romania -- Ministry of Research, Innovation
and Digitization, CNCS-UEFISCDI, contract no.~30N/2023 under Romanian
National Core Program LAPLAS VII, grant no.~PN 23 21 01 02 and project
number PN-III-P1-1.1-TE-2021-0924/TE57/2022, within PNCDI III; Slovenia
-- Slovenian Research Agency, grants P1-0031, P1-0385, I0-0033, N1-0111;
Spain -- Ministerio de Ciencia e Innovaci\'on/Agencia Estatal de
Investigaci\'on (PID2019-105544GB-I00, PID2022-140510NB-I00 and
RYC2019-027017-I), Xunta de Galicia (CIGUS Network of Research Centers,
Consolidaci\'on 2021 GRC GI-2033, ED431C-2021/22 and ED431F-2022/15),
Junta de Andaluc\'\i{}a (SOMM17/6104/UGR and P18-FR-4314), and the European
Union (Marie Sklodowska-Curie 101065027 and ERDF); USA -- Department of
Energy, Contracts No.~DE-AC02-07CH11359, No.~DE-FR02-04ER41300,
No.~DE-FG02-99ER41107 and No.~DE-SC0011689; National Science Foundation,
Grant No.~0450696, and NSF-2013199; The Grainger Foundation; Marie
Curie-IRSES/EPLANET; European Particle Physics Latin American Network;
and UNESCO.
\end{sloppypar}

\bibliographystyle{apsrev4-2}
\bibliography{biblio}

\begin{thebibliography}{50}%
\makeatletter
\providecommand \@ifxundefined [1]{%
 \@ifx{#1\undefined}
}%
\providecommand \@ifnum [1]{%
 \ifnum #1\expandafter \@firstoftwo
 \else \expandafter \@secondoftwo
 \fi
}%
\providecommand \@ifx [1]{%
 \ifx #1\expandafter \@firstoftwo
 \else \expandafter \@secondoftwo
 \fi
}%
\providecommand \natexlab [1]{#1}%
\providecommand \enquote  [1]{``#1''}%
\providecommand \bibnamefont  [1]{#1}%
\providecommand \bibfnamefont [1]{#1}%
\providecommand \citenamefont [1]{#1}%
\providecommand \href@noop [0]{\@secondoftwo}%
\providecommand \href [0]{\begingroup \@sanitize@url \@href}%
\providecommand \@href[1]{\@@startlink{#1}\@@href}%
\providecommand \@@href[1]{\endgroup#1\@@endlink}%
\providecommand \@sanitize@url [0]{\catcode `\\12\catcode `\$12\catcode
  `\&12\catcode `\#12\catcode `\^12\catcode `\_12\catcode `\%12\relax}%
\providecommand \@@startlink[1]{}%
\providecommand \@@endlink[0]{}%
\providecommand \url  [0]{\begingroup\@sanitize@url \@url }%
\providecommand \@url [1]{\endgroup\@href {#1}{\urlprefix }}%
\providecommand \urlprefix  [0]{URL }%
\providecommand \Eprint [0]{\href }%
\providecommand \doibase [0]{https://doi.org/}%
\providecommand \selectlanguage [0]{\@gobble}%
\providecommand \bibinfo  [0]{\@secondoftwo}%
\providecommand \bibfield  [0]{\@secondoftwo}%
\providecommand \translation [1]{[#1]}%
\providecommand \BibitemOpen [0]{}%
\providecommand \bibitemStop [0]{}%
\providecommand \bibitemNoStop [0]{.\EOS\space}%
\providecommand \EOS [0]{\spacefactor3000\relax}%
\providecommand \BibitemShut  [1]{\csname bibitem#1\endcsname}%
\let\auto@bib@innerbib\@empty
\bibitem [{\citenamefont {Aab}\ \emph {et~al.}(2015{\natexlab{a}})\citenamefont
  {Aab} \emph {et~al.}}]{PierreAuger:2015eyc}%
  \BibitemOpen
  \bibfield  {author} {\bibinfo {author} {\bibfnamefont {A.}~\bibnamefont
  {Aab}} \emph {et~al.} (\bibinfo {collaboration} {Pierre Auger}),\ }\href
  {https://doi.org/10.1016/j.nima.2015.06.058} {\bibfield  {journal} {\bibinfo
  {journal} {Nucl. Instrum. Meth. A}\ }\textbf {\bibinfo {volume} {798}},\
  \bibinfo {pages} {172} (\bibinfo {year} {2015}{\natexlab{a}})},\ \Eprint
  {https://arxiv.org/abs/1502.01323} {arXiv:1502.01323 [astro-ph.IM]}
  \BibitemShut {NoStop}%
\bibitem [{\citenamefont {Aab}\ \emph {et~al.}(2017)\citenamefont {Aab} \emph
  {et~al.}}]{PierreAuger:2017pzq}%
  \BibitemOpen
  \bibfield  {author} {\bibinfo {author} {\bibfnamefont {A.}~\bibnamefont
  {Aab}} \emph {et~al.} (\bibinfo {collaboration} {Pierre Auger}),\ }\href
  {https://doi.org/10.1126/science.aan4338} {\bibfield  {journal} {\bibinfo
  {journal} {Science}\ }\textbf {\bibinfo {volume} {357}},\ \bibinfo {pages}
  {1266} (\bibinfo {year} {2017})},\ \Eprint {https://arxiv.org/abs/1709.07321}
  {arXiv:1709.07321 [astro-ph.HE]} \BibitemShut {NoStop}%
\bibitem [{\citenamefont {Halim}\ \emph {et~al.}(2024)\citenamefont {Halim}
  \emph {et~al.}}]{PierreAuger:2024fgl}%
  \BibitemOpen
  \bibfield  {author} {\bibinfo {author} {\bibfnamefont {A.~A.}\ \bibnamefont
  {Halim}} \emph {et~al.} (\bibinfo {collaboration} {Pierre Auger}),\ }\href
  {https://doi.org/10.3847/1538-4357/ad843b} {\bibfield  {journal} {\bibinfo
  {journal} {Astrophys. J.}\ }\textbf {\bibinfo {volume} {976}},\ \bibinfo
  {pages} {48} (\bibinfo {year} {2024})},\ \Eprint
  {https://arxiv.org/abs/2408.05292} {arXiv:2408.05292 [astro-ph.HE]}
  \BibitemShut {NoStop}%
\bibitem [{\citenamefont {Abreu}\ \emph {et~al.}(2022)\citenamefont {Abreu}
  \emph {et~al.}}]{PierreAuger:2022axr}%
  \BibitemOpen
  \bibfield  {author} {\bibinfo {author} {\bibfnamefont {P.}~\bibnamefont
  {Abreu}} \emph {et~al.} (\bibinfo {collaboration} {Pierre Auger}),\ }\href
  {https://doi.org/10.3847/1538-4357/ac7d4e} {\bibfield  {journal} {\bibinfo
  {journal} {Astrophys. J.}\ }\textbf {\bibinfo {volume} {935}},\ \bibinfo
  {pages} {170} (\bibinfo {year} {2022})},\ \Eprint
  {https://arxiv.org/abs/2206.13492} {arXiv:2206.13492 [astro-ph.HE]}
  \BibitemShut {NoStop}%
\bibitem [{\citenamefont {Fujii}(2024)}]{Fujii:2024sys}%
  \BibitemOpen
  \bibfield  {author} {\bibinfo {author} {\bibfnamefont {T.}~\bibnamefont
  {Fujii}},\ }\href {https://doi.org/10.22323/1.444.0031} {\bibfield  {journal}
  {\bibinfo  {journal} {PoS}\ }\textbf {\bibinfo {volume} {ICRC2023}},\
  \bibinfo {pages} {031} (\bibinfo {year} {2024})},\ \Eprint
  {https://arxiv.org/abs/2401.08952} {arXiv:2401.08952 [astro-ph.HE]}
  \BibitemShut {NoStop}%
\bibitem [{\citenamefont {Aab}\ \emph {et~al.}(2020{\natexlab{a}})\citenamefont
  {Aab} \emph {et~al.}}]{PierreAuger:2020qqz}%
  \BibitemOpen
  \bibfield  {author} {\bibinfo {author} {\bibfnamefont {A.}~\bibnamefont
  {Aab}} \emph {et~al.} (\bibinfo {collaboration} {Pierre Auger}),\ }\href
  {https://doi.org/10.1103/PhysRevD.102.062005} {\bibfield  {journal} {\bibinfo
   {journal} {Phys. Rev. D}\ }\textbf {\bibinfo {volume} {102}},\ \bibinfo
  {pages} {062005} (\bibinfo {year} {2020}{\natexlab{a}})},\ \Eprint
  {https://arxiv.org/abs/2008.06486} {arXiv:2008.06486 [astro-ph.HE]}
  \BibitemShut {NoStop}%
\bibitem [{\citenamefont {Aab}\ \emph {et~al.}(2020{\natexlab{b}})\citenamefont
  {Aab} \emph {et~al.}}]{PierreAuger:2020kuy}%
  \BibitemOpen
  \bibfield  {author} {\bibinfo {author} {\bibfnamefont {A.}~\bibnamefont
  {Aab}} \emph {et~al.} (\bibinfo {collaboration} {Pierre Auger}),\ }\href
  {https://doi.org/10.1103/PhysRevLett.125.121106} {\bibfield  {journal}
  {\bibinfo  {journal} {Phys. Rev. Lett.}\ }\textbf {\bibinfo {volume} {125}},\
  \bibinfo {pages} {121106} (\bibinfo {year} {2020}{\natexlab{b}})},\ \Eprint
  {https://arxiv.org/abs/2008.06488} {arXiv:2008.06488 [astro-ph.HE]}
  \BibitemShut {NoStop}%
\bibitem [{\citenamefont {Razzaque}(2020)}]{Razzaque:2020wdo}%
  \BibitemOpen
  \bibfield  {author} {\bibinfo {author} {\bibfnamefont {S.}~\bibnamefont
  {Razzaque}},\ }\href {https://doi.org/10.1103/Physics.13.145} {\bibfield
  {journal} {\bibinfo  {journal} {APS Physics}\ }\textbf {\bibinfo {volume}
  {13}},\ \bibinfo {pages} {145} (\bibinfo {year} {2020})}\BibitemShut
  {NoStop}%
\bibitem [{\citenamefont {Scherb}(1959)}]{Scherb:1959}%
  \BibitemOpen
  \bibfield  {author} {\bibinfo {author} {\bibfnamefont {F.}~\bibnamefont
  {Scherb}},\ }\href@noop {} {\bibfield  {journal} {\bibinfo  {journal} {M.I.T.
  Nuclear Science Technical Report}\ }\textbf {\bibinfo {volume} {71}}
  (\bibinfo {year} {1959})}\BibitemShut {NoStop}%
\bibitem [{\citenamefont {Clark}\ \emph {et~al.}(1961)\citenamefont {Clark},
  \citenamefont {Earl}, \citenamefont {Kraushaar}, \citenamefont {Linsley},
  \citenamefont {Rossi}, \citenamefont {Scherb},\ and\ \citenamefont
  {Scott}}]{Clark:1961mb}%
  \BibitemOpen
  \bibfield  {author} {\bibinfo {author} {\bibfnamefont {G.~W.}\ \bibnamefont
  {Clark}}, \bibinfo {author} {\bibfnamefont {J.}~\bibnamefont {Earl}},
  \bibinfo {author} {\bibfnamefont {W.~L.}\ \bibnamefont {Kraushaar}}, \bibinfo
  {author} {\bibfnamefont {J.}~\bibnamefont {Linsley}}, \bibinfo {author}
  {\bibfnamefont {B.~B.}\ \bibnamefont {Rossi}}, \bibinfo {author}
  {\bibfnamefont {F.}~\bibnamefont {Scherb}},\ and\ \bibinfo {author}
  {\bibfnamefont {D.~W.}\ \bibnamefont {Scott}},\ }\href
  {https://doi.org/10.1103/PhysRev.122.637} {\bibfield  {journal} {\bibinfo
  {journal} {Phys. Rev.}\ }\textbf {\bibinfo {volume} {122}},\ \bibinfo {pages}
  {637} (\bibinfo {year} {1961})}\BibitemShut {NoStop}%
\bibitem [{\citenamefont {Ave}\ \emph {et~al.}(2000{\natexlab{a}})\citenamefont
  {Ave}, \citenamefont {Hinton}, \citenamefont {Vazquez}, \citenamefont
  {Watson},\ and\ \citenamefont {Zas}}]{Ave:2000nd}%
  \BibitemOpen
  \bibfield  {author} {\bibinfo {author} {\bibfnamefont {M.}~\bibnamefont
  {Ave}}, \bibinfo {author} {\bibfnamefont {J.~A.}\ \bibnamefont {Hinton}},
  \bibinfo {author} {\bibfnamefont {R.~A.}\ \bibnamefont {Vazquez}}, \bibinfo
  {author} {\bibfnamefont {A.~A.}\ \bibnamefont {Watson}},\ and\ \bibinfo
  {author} {\bibfnamefont {E.}~\bibnamefont {Zas}},\ }\href
  {https://doi.org/10.1103/PhysRevLett.85.2244} {\bibfield  {journal} {\bibinfo
   {journal} {Phys. Rev. Lett.}\ }\textbf {\bibinfo {volume} {85}},\ \bibinfo
  {pages} {2244} (\bibinfo {year} {2000}{\natexlab{a}})},\ \Eprint
  {https://arxiv.org/abs/astro-ph/0007386} {arXiv:astro-ph/0007386}
  \BibitemShut {NoStop}%
\bibitem [{\citenamefont {Aab}\ \emph {et~al.}(2020{\natexlab{c}})\citenamefont
  {Aab} \emph {et~al.}}]{PierreAuger:2020yab}%
  \BibitemOpen
  \bibfield  {author} {\bibinfo {author} {\bibfnamefont {A.}~\bibnamefont
  {Aab}} \emph {et~al.} (\bibinfo {collaboration} {Pierre Auger}),\ }\href
  {https://doi.org/10.1088/1748-0221/15/10/P10021} {\bibfield  {journal}
  {\bibinfo  {journal} {JINST}\ }\textbf {\bibinfo {volume} {15}}\bibfield
  {number} {\bibinfo  {number} { (10)},\ \bibinfo {pages} {P10021}},\ }\Eprint
  {https://arxiv.org/abs/2007.09035} {arXiv:2007.09035 [astro-ph.IM]}
  \BibitemShut {NoStop}%
\bibitem [{\citenamefont {Ave}\ \emph {et~al.}(2000{\natexlab{b}})\citenamefont
  {Ave}, \citenamefont {Vazquez},\ and\ \citenamefont {Zas}}]{Ave:2000xs}%
  \BibitemOpen
  \bibfield  {author} {\bibinfo {author} {\bibfnamefont {M.}~\bibnamefont
  {Ave}}, \bibinfo {author} {\bibfnamefont {R.~A.}\ \bibnamefont {Vazquez}},\
  and\ \bibinfo {author} {\bibfnamefont {E.}~\bibnamefont {Zas}},\ }\href
  {https://doi.org/10.1016/S0927-6505(00)00113-4} {\bibfield  {journal}
  {\bibinfo  {journal} {Astropart. Phys.}\ }\textbf {\bibinfo {volume} {14}},\
  \bibinfo {pages} {91} (\bibinfo {year} {2000}{\natexlab{b}})},\ \Eprint
  {https://arxiv.org/abs/astro-ph/0011490} {arXiv:astro-ph/0011490}
  \BibitemShut {NoStop}%
\bibitem [{\citenamefont {Aab}\ \emph {et~al.}(2014{\natexlab{a}})\citenamefont
  {Aab} \emph {et~al.}}]{PierreAuger:2014jss}%
  \BibitemOpen
  \bibfield  {author} {\bibinfo {author} {\bibfnamefont {A.}~\bibnamefont
  {Aab}} \emph {et~al.} (\bibinfo {collaboration} {Pierre Auger}),\ }\href
  {https://doi.org/10.1088/1475-7516/2014/08/019} {\bibfield  {journal}
  {\bibinfo  {journal} {JCAP}\ }\textbf {\bibinfo {volume} {08}},\ \bibinfo
  {pages} {019 (2014)}},\ \Eprint {https://arxiv.org/abs/1407.3214}
  {arXiv:1407.3214 [astro-ph.HE]} \BibitemShut {NoStop}%
\bibitem [{\citenamefont {Hersil}\ \emph {et~al.}(1961)\citenamefont {Hersil},
  \citenamefont {Escobar}, \citenamefont {Scott}, \citenamefont {Clark},\ and\
  \citenamefont {Olbert}}]{Hersil:1961zz}%
  \BibitemOpen
  \bibfield  {author} {\bibinfo {author} {\bibfnamefont {J.}~\bibnamefont
  {Hersil}}, \bibinfo {author} {\bibfnamefont {I.}~\bibnamefont {Escobar}},
  \bibinfo {author} {\bibfnamefont {D.}~\bibnamefont {Scott}}, \bibinfo
  {author} {\bibfnamefont {G.}~\bibnamefont {Clark}},\ and\ \bibinfo {author}
  {\bibfnamefont {S.}~\bibnamefont {Olbert}},\ }\href
  {https://doi.org/10.1103/PhysRevLett.6.22} {\bibfield  {journal} {\bibinfo
  {journal} {Phys. Rev. Lett.}\ }\textbf {\bibinfo {volume} {6}},\ \bibinfo
  {pages} {22} (\bibinfo {year} {1961})}\BibitemShut {NoStop}%
\bibitem [{\citenamefont {Aab}\ \emph {et~al.}(2019)\citenamefont {Aab} \emph
  {et~al.}}]{PierreAuger:2019dhr}%
  \BibitemOpen
  \bibfield  {author} {\bibinfo {author} {\bibfnamefont {A.}~\bibnamefont
  {Aab}} \emph {et~al.} (\bibinfo {collaboration} {Pierre Auger}),\ }\href
  {https://doi.org/10.1103/PhysRevD.100.082003} {\bibfield  {journal} {\bibinfo
   {journal} {Phys. Rev. D}\ }\textbf {\bibinfo {volume} {100}},\ \bibinfo
  {pages} {082003} (\bibinfo {year} {2019})},\ \Eprint
  {https://arxiv.org/abs/1901.08040} {arXiv:1901.08040 [astro-ph.IM]}
  \BibitemShut {NoStop}%
\bibitem [{\citenamefont {Aab}\ \emph {et~al.}(2015{\natexlab{b}})\citenamefont
  {Aab} \emph {et~al.}}]{PierreAuger:2015xho}%
  \BibitemOpen
  \bibfield  {author} {\bibinfo {author} {\bibfnamefont {A.}~\bibnamefont
  {Aab}} \emph {et~al.} (\bibinfo {collaboration} {Pierre Auger}),\ }\href
  {https://doi.org/10.1088/1475-7516/2015/08/049} {\bibfield  {journal}
  {\bibinfo  {journal} {JCAP}\ }\textbf {\bibinfo {volume} {08}},\ \bibinfo
  {pages} {049 (2015)}},\ \Eprint {https://arxiv.org/abs/1503.07786}
  {arXiv:1503.07786 [astro-ph.HE]} \BibitemShut {NoStop}%
\bibitem [{\citenamefont {Dembinski}\ \emph {et~al.}(2010)\citenamefont
  {Dembinski}, \citenamefont {Billoir}, \citenamefont {Deligny},\ and\
  \citenamefont {Hebbeker}}]{Dembinski:2009jc}%
  \BibitemOpen
  \bibfield  {author} {\bibinfo {author} {\bibfnamefont {H.~P.}\ \bibnamefont
  {Dembinski}}, \bibinfo {author} {\bibfnamefont {P.}~\bibnamefont {Billoir}},
  \bibinfo {author} {\bibfnamefont {O.}~\bibnamefont {Deligny}},\ and\ \bibinfo
  {author} {\bibfnamefont {T.}~\bibnamefont {Hebbeker}},\ }\href
  {https://doi.org/10.1016/j.astropartphys.2010.06.006} {\bibfield  {journal}
  {\bibinfo  {journal} {Astropart. Phys.}\ }\textbf {\bibinfo {volume} {34}},\
  \bibinfo {pages} {128} (\bibinfo {year} {2010})},\ \Eprint
  {https://arxiv.org/abs/0904.2372} {arXiv:0904.2372 [astro-ph.IM]}
  \BibitemShut {NoStop}%
\bibitem [{\citenamefont {Abreu}\ \emph {et~al.}(2021)\citenamefont {Abreu}
  \emph {et~al.}}]{PierreAuger:2021hun}%
  \BibitemOpen
  \bibfield  {author} {\bibinfo {author} {\bibfnamefont {P.}~\bibnamefont
  {Abreu}} \emph {et~al.} (\bibinfo {collaboration} {Pierre Auger}),\ }\href
  {https://doi.org/10.1140/epjc/s10052-021-09700-w} {\bibfield  {journal}
  {\bibinfo  {journal} {Eur. Phys. J. C}\ }\textbf {\bibinfo {volume} {81}},\
  \bibinfo {pages} {966} (\bibinfo {year} {2021})},\ \Eprint
  {https://arxiv.org/abs/2109.13400} {arXiv:2109.13400 [astro-ph.HE]}
  \BibitemShut {NoStop}%
\bibitem [{\citenamefont {Verzi}\ \emph {et~al.}(2013)\citenamefont {Verzi}
  \emph {et~al.}}]{Verzi:2013ajy}%
  \BibitemOpen
  \bibfield  {author} {\bibinfo {author} {\bibfnamefont {V.}~\bibnamefont
  {Verzi}} \emph {et~al.} (\bibinfo {collaboration} {Pierre Auger}),\ }in\
  \href@noop {} {\emph {\bibinfo {booktitle} {{Proceedings of 33rd ICRC}}}}\
  (\bibinfo {year} {2013})\ p.\ \bibinfo {pages} {0928}\BibitemShut {NoStop}%
\bibitem [{\citenamefont {Abraham}\ \emph {et~al.}(2010)\citenamefont {Abraham}
  \emph {et~al.}}]{PierreAuger:2010zof}%
  \BibitemOpen
  \bibfield  {author} {\bibinfo {author} {\bibfnamefont {J.}~\bibnamefont
  {Abraham}} \emph {et~al.} (\bibinfo {collaboration} {Pierre Auger}),\ }\href
  {https://doi.org/10.1016/j.nima.2009.11.018} {\bibfield  {journal} {\bibinfo
  {journal} {Nucl. Instrum. Meth. A}\ }\textbf {\bibinfo {volume} {613}},\
  \bibinfo {pages} {29} (\bibinfo {year} {2010})},\ \Eprint
  {https://arxiv.org/abs/1111.6764} {arXiv:1111.6764 [astro-ph.IM]}
  \BibitemShut {NoStop}%
\bibitem [{\citenamefont {Guerard}\ \emph {et~al.}(2002)\citenamefont {Guerard}
  \emph {et~al.}}]{Guerard:2002yg}%
  \BibitemOpen
  \bibfield  {author} {\bibinfo {author} {\bibfnamefont {C.~K.}\ \bibnamefont
  {Guerard}} \emph {et~al.} (\bibinfo {collaboration} {Pierre Auger}),\ }in\
  \href@noop {} {\emph {\bibinfo {booktitle} {{Proceedings of 27th ICRC}}}}\
  (\bibinfo {year} {2002})\BibitemShut {NoStop}%
\bibitem [{\citenamefont {Abreu}\ \emph {et~al.}(2011)\citenamefont {Abreu}
  \emph {et~al.}}]{PierreAuger:2010swb}%
  \BibitemOpen
  \bibfield  {author} {\bibinfo {author} {\bibfnamefont {P.}~\bibnamefont
  {Abreu}} \emph {et~al.} (\bibinfo {collaboration} {Pierre Auger}),\ }\href
  {https://doi.org/10.1016/j.astropartphys.2010.10.001} {\bibfield  {journal}
  {\bibinfo  {journal} {Astropart. Phys.}\ }\textbf {\bibinfo {volume} {34}},\
  \bibinfo {pages} {368} (\bibinfo {year} {2011})},\ \Eprint
  {https://arxiv.org/abs/1010.6162} {arXiv:1010.6162 [astro-ph.HE]}
  \BibitemShut {NoStop}%
\bibitem [{\citenamefont {Abdul~Halim}\ \emph
  {et~al.}(2025{\natexlab{a}})\citenamefont {Abdul~Halim} \emph
  {et~al.}}]{PierreAuger:2024flk}%
  \BibitemOpen
  \bibfield  {author} {\bibinfo {author} {\bibfnamefont {A.}~\bibnamefont
  {Abdul~Halim}} \emph {et~al.} (\bibinfo {collaboration} {Pierre Auger}),\
  }\href {https://doi.org/10.1103/PhysRevLett.134.021001} {\bibfield  {journal}
  {\bibinfo  {journal} {Phys. Rev. Lett.}\ }\textbf {\bibinfo {volume} {134}},\
  \bibinfo {pages} {021001} (\bibinfo {year} {2025}{\natexlab{a}})},\ \Eprint
  {https://arxiv.org/abs/2406.06315} {arXiv:2406.06315 [astro-ph.HE]}
  \BibitemShut {NoStop}%
\bibitem [{\citenamefont {Abdul~Halim}\ \emph
  {et~al.}(2025{\natexlab{b}})\citenamefont {Abdul~Halim} \emph
  {et~al.}}]{PierreAuger:2024nzw}%
  \BibitemOpen
  \bibfield  {author} {\bibinfo {author} {\bibfnamefont {A.}~\bibnamefont
  {Abdul~Halim}} \emph {et~al.} (\bibinfo {collaboration} {Pierre Auger}),\
  }\href {https://doi.org/10.1103/PhysRevD.111.022003} {\bibfield  {journal}
  {\bibinfo  {journal} {Phys. Rev. D}\ }\textbf {\bibinfo {volume} {111}},\
  \bibinfo {pages} {022003} (\bibinfo {year} {2025}{\natexlab{b}})},\ \Eprint
  {https://arxiv.org/abs/2406.06319} {arXiv:2406.06319 [astro-ph.HE]}
  \BibitemShut {NoStop}%
\bibitem [{\citenamefont {Bonamente}(2020)}]{Bonamente:2019efn}%
  \BibitemOpen
  \bibfield  {author} {\bibinfo {author} {\bibfnamefont {M.}~\bibnamefont
  {Bonamente}},\ }\href {https://doi.org/10.1080/02664763.2019.1704703}
  {\bibfield  {journal} {\bibinfo  {journal} {J. Appl. Stat.}\ }\textbf
  {\bibinfo {volume} {47}},\ \bibinfo {pages} {2044} (\bibinfo {year}
  {2020})},\ \Eprint {https://arxiv.org/abs/1912.05444} {arXiv:1912.05444
  [astro-ph.HE]} \BibitemShut {NoStop}%
\bibitem [{\citenamefont {Abreu}\ \emph {et~al.}(2012)\citenamefont {Abreu}
  \emph {et~al.}}]{PierreAuger:2012xju}%
  \BibitemOpen
  \bibfield  {author} {\bibinfo {author} {\bibfnamefont {P.}~\bibnamefont
  {Abreu}} \emph {et~al.} (\bibinfo {collaboration} {Pierre Auger}),\ }\href
  {https://doi.org/10.1088/0067-0049/203/2/34} {\bibfield  {journal} {\bibinfo
  {journal} {Astrophys. J. Suppl.}\ }\textbf {\bibinfo {volume} {203}},\
  \bibinfo {pages} {34} (\bibinfo {year} {2012})},\ \Eprint
  {https://arxiv.org/abs/1210.3736} {arXiv:1210.3736 [astro-ph.HE]}
  \BibitemShut {NoStop}%
\bibitem [{\citenamefont {Biermann}\ and\ \citenamefont
  {de~Souza}(2012)}]{Biermann:2011wf}%
  \BibitemOpen
  \bibfield  {author} {\bibinfo {author} {\bibfnamefont {P.~L.}\ \bibnamefont
  {Biermann}}\ and\ \bibinfo {author} {\bibfnamefont {V.}~\bibnamefont
  {de~Souza}},\ }\href {https://doi.org/10.1088/0004-637X/746/1/72} {\bibfield
  {journal} {\bibinfo  {journal} {Astrophys. J.}\ }\textbf {\bibinfo {volume}
  {746}},\ \bibinfo {pages} {72} (\bibinfo {year} {2012})},\ \Eprint
  {https://arxiv.org/abs/1106.0625} {arXiv:1106.0625 [astro-ph.HE]}
  \BibitemShut {NoStop}%
\bibitem [{\citenamefont {Keivani}\ \emph {et~al.}(2014)\citenamefont
  {Keivani}, \citenamefont {Farrar},\ and\ \citenamefont
  {Sutherland}}]{Keivani:2014kua}%
  \BibitemOpen
  \bibfield  {author} {\bibinfo {author} {\bibfnamefont {A.}~\bibnamefont
  {Keivani}}, \bibinfo {author} {\bibfnamefont {G.~R.}\ \bibnamefont
  {Farrar}},\ and\ \bibinfo {author} {\bibfnamefont {M.}~\bibnamefont
  {Sutherland}},\ }\href {https://doi.org/10.1016/j.astropartphys.2014.07.001}
  {\bibfield  {journal} {\bibinfo  {journal} {Astropart. Phys.}\ }\textbf
  {\bibinfo {volume} {61}},\ \bibinfo {pages} {47} (\bibinfo {year} {2014})},\
  \Eprint {https://arxiv.org/abs/1406.5249} {arXiv:1406.5249 [astro-ph.HE]}
  \BibitemShut {NoStop}%
\bibitem [{\citenamefont {Bell}\ and\ \citenamefont
  {Matthews}(2022)}]{Bell:2021pkk}%
  \BibitemOpen
  \bibfield  {author} {\bibinfo {author} {\bibfnamefont {A.~R.}\ \bibnamefont
  {Bell}}\ and\ \bibinfo {author} {\bibfnamefont {J.~H.}\ \bibnamefont
  {Matthews}},\ }\href {https://doi.org/10.1093/mnras/stac031} {\bibfield
  {journal} {\bibinfo  {journal} {Mon. Not. Roy. Astron. Soc.}\ }\textbf
  {\bibinfo {volume} {511}},\ \bibinfo {pages} {448} (\bibinfo {year}
  {2022})},\ \Eprint {https://arxiv.org/abs/2108.08879} {arXiv:2108.08879
  [astro-ph.HE]} \BibitemShut {NoStop}%
\bibitem [{\citenamefont {Mollerach}\ and\ \citenamefont
  {Roulet}(2024)}]{Mollerach:2024sjd}%
  \BibitemOpen
  \bibfield  {author} {\bibinfo {author} {\bibfnamefont {S.}~\bibnamefont
  {Mollerach}}\ and\ \bibinfo {author} {\bibfnamefont {E.}~\bibnamefont
  {Roulet}},\ }\href {https://doi.org/10.1103/PhysRevD.110.063030} {\bibfield
  {journal} {\bibinfo  {journal} {Phys. Rev. D}\ }\textbf {\bibinfo {volume}
  {110}},\ \bibinfo {pages} {063030} (\bibinfo {year} {2024})},\ \Eprint
  {https://arxiv.org/abs/2406.19199} {arXiv:2406.19199 [astro-ph.HE]}
  \BibitemShut {NoStop}%
\bibitem [{\citenamefont {Marafico}\ \emph {et~al.}(2024)\citenamefont
  {Marafico}, \citenamefont {Biteau}, \citenamefont {Condorelli}, \citenamefont
  {Deligny},\ and\ \citenamefont {Bregeon}}]{Marafico:2024qgh}%
  \BibitemOpen
  \bibfield  {author} {\bibinfo {author} {\bibfnamefont {S.}~\bibnamefont
  {Marafico}}, \bibinfo {author} {\bibfnamefont {J.}~\bibnamefont {Biteau}},
  \bibinfo {author} {\bibfnamefont {A.}~\bibnamefont {Condorelli}}, \bibinfo
  {author} {\bibfnamefont {O.}~\bibnamefont {Deligny}},\ and\ \bibinfo {author}
  {\bibfnamefont {J.}~\bibnamefont {Bregeon}},\ }\href
  {https://doi.org/10.3847/1538-4357/ad5a11} {\bibfield  {journal} {\bibinfo
  {journal} {Astrophys. J.}\ }\textbf {\bibinfo {volume} {972}},\ \bibinfo
  {pages} {4} (\bibinfo {year} {2024})},\ \Eprint
  {https://arxiv.org/abs/2405.17179} {arXiv:2405.17179 [astro-ph.HE]}
  \BibitemShut {NoStop}%
\bibitem [{\citenamefont {Mbarek}\ \emph {et~al.}(2025)\citenamefont {Mbarek},
  \citenamefont {Caprioli},\ and\ \citenamefont {Murase}}]{Mbarek:2024nvv}%
  \BibitemOpen
  \bibfield  {author} {\bibinfo {author} {\bibfnamefont {R.}~\bibnamefont
  {Mbarek}}, \bibinfo {author} {\bibfnamefont {D.}~\bibnamefont {Caprioli}},\
  and\ \bibinfo {author} {\bibfnamefont {K.}~\bibnamefont {Murase}},\ }\href
  {https://doi.org/10.1103/PhysRevD.111.023024} {\bibfield  {journal} {\bibinfo
   {journal} {Phys. Rev. D}\ }\textbf {\bibinfo {volume} {111}},\ \bibinfo
  {pages} {023024} (\bibinfo {year} {2025})},\ \Eprint
  {https://arxiv.org/abs/2410.05696} {arXiv:2410.05696 [astro-ph.HE]}
  \BibitemShut {NoStop}%
\bibitem [{\citenamefont {Romero}\ \emph {et~al.}(2018)\citenamefont {Romero},
  \citenamefont {M\"uller},\ and\ \citenamefont {Roth}}]{Romero:2018mnb}%
  \BibitemOpen
  \bibfield  {author} {\bibinfo {author} {\bibfnamefont {G.~E.}\ \bibnamefont
  {Romero}}, \bibinfo {author} {\bibfnamefont {A.~L.}\ \bibnamefont
  {M\"uller}},\ and\ \bibinfo {author} {\bibfnamefont {M.}~\bibnamefont
  {Roth}},\ }\href {https://doi.org/10.1051/0004-6361/201832666} {\bibfield
  {journal} {\bibinfo  {journal} {Astron. Astrophys.}\ }\textbf {\bibinfo
  {volume} {616}},\ \bibinfo {pages} {A57} (\bibinfo {year} {2018})},\ \Eprint
  {https://arxiv.org/abs/1801.06483} {arXiv:1801.06483 [astro-ph.HE]}
  \BibitemShut {NoStop}%
\bibitem [{\citenamefont {Anchordoqui}\ and\ \citenamefont
  {Soriano}(2021)}]{Anchordoqui:2019mfb}%
  \BibitemOpen
  \bibfield  {author} {\bibinfo {author} {\bibfnamefont {L.~A.}\ \bibnamefont
  {Anchordoqui}}\ and\ \bibinfo {author} {\bibfnamefont {J.~F.}\ \bibnamefont
  {Soriano}},\ }\href {https://doi.org/10.22323/1.358.0255} {\bibfield
  {journal} {\bibinfo  {journal} {PoS}\ }\textbf {\bibinfo {volume}
  {ICRC2019}},\ \bibinfo {pages} {255} (\bibinfo {year} {2021})},\ \Eprint
  {https://arxiv.org/abs/1905.13243} {arXiv:1905.13243 [astro-ph.HE]}
  \BibitemShut {NoStop}%
\bibitem [{\citenamefont {Abbasi}\ \emph {et~al.}(2024)\citenamefont {Abbasi}
  \emph {et~al.}}]{TelescopeArray:2024tbi}%
  \BibitemOpen
  \bibfield  {author} {\bibinfo {author} {\bibfnamefont {R.~U.}\ \bibnamefont
  {Abbasi}} \emph {et~al.} (\bibinfo {collaboration} {Telescope Array}),\
  }\href@noop {} {\  (\bibinfo {year} {2024})},\ \Eprint
  {https://arxiv.org/abs/2406.08612} {arXiv:2406.08612 [astro-ph.HE]}
  \BibitemShut {NoStop}%
\bibitem [{\citenamefont {Halim}\ \emph {et~al.}(2023)\citenamefont {Halim}
  \emph {et~al.}}]{PierreAuger:2022atd}%
  \BibitemOpen
  \bibfield  {author} {\bibinfo {author} {\bibfnamefont {A.~A.}\ \bibnamefont
  {Halim}} \emph {et~al.} (\bibinfo {collaboration} {Pierre Auger}),\ }\href
  {https://doi.org/10.1088/1475-7516/2023/05/024} {\bibfield  {journal}
  {\bibinfo  {journal} {JCAP}\ }\textbf {\bibinfo {volume} {05}},\ \bibinfo
  {pages} {024}},\ \Eprint {https://arxiv.org/abs/2211.02857} {arXiv:2211.02857
  [astro-ph.HE]} \BibitemShut {NoStop}%
\bibitem [{\citenamefont {Ehlert}\ \emph {et~al.}(2023)\citenamefont {Ehlert},
  \citenamefont {Oikonomou},\ and\ \citenamefont {Unger}}]{Ehlert:2022jmy}%
  \BibitemOpen
  \bibfield  {author} {\bibinfo {author} {\bibfnamefont {D.}~\bibnamefont
  {Ehlert}}, \bibinfo {author} {\bibfnamefont {F.}~\bibnamefont {Oikonomou}},\
  and\ \bibinfo {author} {\bibfnamefont {M.}~\bibnamefont {Unger}},\ }\href
  {https://doi.org/10.1103/PhysRevD.107.103045} {\bibfield  {journal} {\bibinfo
   {journal} {Phys. Rev. D}\ }\textbf {\bibinfo {volume} {107}},\ \bibinfo
  {pages} {103045} (\bibinfo {year} {2023})},\ \Eprint
  {https://arxiv.org/abs/2207.10691} {arXiv:2207.10691 [astro-ph.HE]}
  \BibitemShut {NoStop}%
\bibitem [{\citenamefont {Farrar}(2025)}]{Farrar:2024zsm}%
  \BibitemOpen
  \bibfield  {author} {\bibinfo {author} {\bibfnamefont {G.~R.}\ \bibnamefont
  {Farrar}},\ }\href {https://doi.org/10.1103/PhysRevLett.134.081003}
  {\bibfield  {journal} {\bibinfo  {journal} {Phys. Rev. Lett.}\ }\textbf
  {\bibinfo {volume} {134}},\ \bibinfo {pages} {081003} (\bibinfo {year}
  {2025})},\ \Eprint {https://arxiv.org/abs/2405.12004} {arXiv:2405.12004
  [astro-ph.HE]} \BibitemShut {NoStop}%
\bibitem [{\citenamefont {Zhang}\ \emph {et~al.}(2024)\citenamefont {Zhang},
  \citenamefont {Murase}, \citenamefont {Ekanger}, \citenamefont
  {Bhattacharya},\ and\ \citenamefont {Horiuchi}}]{Zhang:2024sjp}%
  \BibitemOpen
  \bibfield  {author} {\bibinfo {author} {\bibfnamefont {B.~T.}\ \bibnamefont
  {Zhang}}, \bibinfo {author} {\bibfnamefont {K.}~\bibnamefont {Murase}},
  \bibinfo {author} {\bibfnamefont {N.}~\bibnamefont {Ekanger}}, \bibinfo
  {author} {\bibfnamefont {M.}~\bibnamefont {Bhattacharya}},\ and\ \bibinfo
  {author} {\bibfnamefont {S.}~\bibnamefont {Horiuchi}},\ }\href@noop {} {\
  (\bibinfo {year} {2024})},\ \Eprint {https://arxiv.org/abs/2405.17409}
  {arXiv:2405.17409 [astro-ph.HE]} \BibitemShut {NoStop}%
\bibitem [{\citenamefont {Aab}\ \emph {et~al.}(2016)\citenamefont {Aab} \emph
  {et~al.}}]{PierreAuger:2016qzd}%
  \BibitemOpen
  \bibfield  {author} {\bibinfo {author} {\bibfnamefont {A.}~\bibnamefont
  {Aab}} \emph {et~al.} (\bibinfo {collaboration} {Pierre Auger}),\ }\href@noop
  {} {\  (\bibinfo {year} {2016})},\ \Eprint {https://arxiv.org/abs/1604.03637}
  {arXiv:1604.03637 [astro-ph.IM]} \BibitemShut {NoStop}%
\bibitem [{\citenamefont {Dawson}\ \emph {et~al.}(2019)\citenamefont {Dawson}
  \emph {et~al.}}]{Dawson:2019zva}%
  \BibitemOpen
  \bibfield  {author} {\bibinfo {author} {\bibfnamefont {B.~R.}\ \bibnamefont
  {Dawson}} \emph {et~al.} (\bibinfo {collaboration} {Pierre Auger}),\ }\href
  {https://doi.org/10.1051/epjconf/201919701004} {\bibfield  {journal}
  {\bibinfo  {journal} {EPJ Web Conf.}\ }\textbf {\bibinfo {volume} {197}},\
  \bibinfo {pages} {01004} (\bibinfo {year} {2019})}\BibitemShut {NoStop}%
\bibitem [{\citenamefont {Harvey}\ \emph {et~al.}(2023)\citenamefont {Harvey}
  \emph {et~al.}}]{PierreAuger:2023nbk}%
  \BibitemOpen
  \bibfield  {author} {\bibinfo {author} {\bibfnamefont {V.~M.}\ \bibnamefont
  {Harvey}} \emph {et~al.} (\bibinfo {collaboration} {Pierre Auger}),\ }\href
  {https://doi.org/10.22323/1.444.0300} {\bibfield  {journal} {\bibinfo
  {journal} {PoS}\ }\textbf {\bibinfo {volume} {ICRC2023}},\ \bibinfo {pages}
  {300} (\bibinfo {year} {2023})}\BibitemShut {NoStop}%
\bibitem [{\citenamefont {Bellido}\ \emph {et~al.}(2023)\citenamefont {Bellido}
  \emph {et~al.}}]{PierreAuger:2023att}%
  \BibitemOpen
  \bibfield  {author} {\bibinfo {author} {\bibfnamefont {J.}~\bibnamefont
  {Bellido}} \emph {et~al.} (\bibinfo {collaboration} {Pierre Auger}),\ }\href
  {https://doi.org/10.22323/1.444.0211} {\bibfield  {journal} {\bibinfo
  {journal} {PoS}\ }\textbf {\bibinfo {volume} {ICRC2023}},\ \bibinfo {pages}
  {211} (\bibinfo {year} {2023})}\BibitemShut {NoStop}%
\bibitem [{\citenamefont {Aab}\ \emph {et~al.}(2014{\natexlab{b}})\citenamefont
  {Aab} \emph {et~al.}}]{PierreAuger:2014sui}%
  \BibitemOpen
  \bibfield  {author} {\bibinfo {author} {\bibfnamefont {A.}~\bibnamefont
  {Aab}} \emph {et~al.} (\bibinfo {collaboration} {Pierre Auger}),\ }\href
  {https://doi.org/10.1103/PhysRevD.90.122005} {\bibfield  {journal} {\bibinfo
  {journal} {Phys. Rev. D}\ }\textbf {\bibinfo {volume} {90}},\ \bibinfo
  {pages} {122005} (\bibinfo {year} {2014}{\natexlab{b}})},\ \Eprint
  {https://arxiv.org/abs/1409.4809} {arXiv:1409.4809 [astro-ph.HE]}
  \BibitemShut {NoStop}%
\bibitem [{\citenamefont {Matthews}(2005)}]{Matthews:2005sd}%
  \BibitemOpen
  \bibfield  {author} {\bibinfo {author} {\bibfnamefont {J.}~\bibnamefont
  {Matthews}},\ }\href {https://doi.org/10.1016/j.astropartphys.2004.09.003}
  {\bibfield  {journal} {\bibinfo  {journal} {Astropart. Phys.}\ }\textbf
  {\bibinfo {volume} {22}},\ \bibinfo {pages} {387} (\bibinfo {year}
  {2005})}\BibitemShut {NoStop}%
\bibitem [{\citenamefont {Pierog}\ \emph {et~al.}(2015)\citenamefont {Pierog},
  \citenamefont {Karpenko}, \citenamefont {Katzy}, \citenamefont {Yatsenko},\
  and\ \citenamefont {Werner}}]{Pierog:2013ria}%
  \BibitemOpen
  \bibfield  {author} {\bibinfo {author} {\bibfnamefont {T.}~\bibnamefont
  {Pierog}}, \bibinfo {author} {\bibfnamefont {I.}~\bibnamefont {Karpenko}},
  \bibinfo {author} {\bibfnamefont {J.~M.}\ \bibnamefont {Katzy}}, \bibinfo
  {author} {\bibfnamefont {E.}~\bibnamefont {Yatsenko}},\ and\ \bibinfo
  {author} {\bibfnamefont {K.}~\bibnamefont {Werner}},\ }\href
  {https://doi.org/10.1103/PhysRevC.92.034906} {\bibfield  {journal} {\bibinfo
  {journal} {Phys. Rev. C}\ }\textbf {\bibinfo {volume} {92}},\ \bibinfo
  {pages} {034906} (\bibinfo {year} {2015})},\ \Eprint
  {https://arxiv.org/abs/1306.0121} {arXiv:1306.0121 [hep-ph]} \BibitemShut
  {NoStop}%
\bibitem [{\citenamefont {Cazon}\ \emph {et~al.}(2018)\citenamefont {Cazon},
  \citenamefont {Concei\c{c}\~ao},\ and\ \citenamefont
  {Riehn}}]{Cazon:2018gww}%
  \BibitemOpen
  \bibfield  {author} {\bibinfo {author} {\bibfnamefont {L.}~\bibnamefont
  {Cazon}}, \bibinfo {author} {\bibfnamefont {R.}~\bibnamefont
  {Concei\c{c}\~ao}},\ and\ \bibinfo {author} {\bibfnamefont {F.}~\bibnamefont
  {Riehn}},\ }\href {https://doi.org/10.1016/j.physletb.2018.07.026} {\bibfield
   {journal} {\bibinfo  {journal} {Phys. Lett. B}\ }\textbf {\bibinfo {volume}
  {784}},\ \bibinfo {pages} {68} (\bibinfo {year} {2018})},\ \Eprint
  {https://arxiv.org/abs/1803.05699} {arXiv:1803.05699 [hep-ph]} \BibitemShut
  {NoStop}%
\bibitem [{\citenamefont {B{\'e}rat}\ \emph {et~al.}(2022)\citenamefont
  {B{\'e}rat}, \citenamefont {Bleve}, \citenamefont {Deligny}, \citenamefont
  {Montanet}, \citenamefont {Savina},\ and\ \citenamefont
  {Torr{\`e}s}}]{Berat:2022iea}%
  \BibitemOpen
  \bibfield  {author} {\bibinfo {author} {\bibfnamefont {C.}~\bibnamefont
  {B{\'e}rat}}, \bibinfo {author} {\bibfnamefont {C.}~\bibnamefont {Bleve}},
  \bibinfo {author} {\bibfnamefont {O.}~\bibnamefont {Deligny}}, \bibinfo
  {author} {\bibfnamefont {F.}~\bibnamefont {Montanet}}, \bibinfo {author}
  {\bibfnamefont {P.}~\bibnamefont {Savina}},\ and\ \bibinfo {author}
  {\bibfnamefont {Z.}~\bibnamefont {Torr{\`e}s}},\ }\href
  {https://doi.org/10.3847/1538-4357/ac5cbe} {\bibfield  {journal} {\bibinfo
  {journal} {Astrophys. J.}\ }\textbf {\bibinfo {volume} {929}},\ \bibinfo
  {pages} {55} (\bibinfo {year} {2022})},\ \Eprint
  {https://arxiv.org/abs/2203.08751} {arXiv:2203.08751 [astro-ph.HE]}
  \BibitemShut {NoStop}%
\bibitem [{\citenamefont {Berat}\ \emph {et~al.}(2024)\citenamefont {Berat},
  \citenamefont {Condorelli}, \citenamefont {Deligny}, \citenamefont
  {Montanet},\ and\ \citenamefont {Torres}}]{Berat:2024rvf}%
  \BibitemOpen
  \bibfield  {author} {\bibinfo {author} {\bibfnamefont {C.}~\bibnamefont
  {Berat}}, \bibinfo {author} {\bibfnamefont {A.}~\bibnamefont {Condorelli}},
  \bibinfo {author} {\bibfnamefont {O.}~\bibnamefont {Deligny}}, \bibinfo
  {author} {\bibfnamefont {F.}~\bibnamefont {Montanet}},\ and\ \bibinfo
  {author} {\bibfnamefont {Z.}~\bibnamefont {Torres}},\ }\href
  {https://doi.org/10.3847/1538-4357/ad372a} {\bibfield  {journal} {\bibinfo
  {journal} {Astrophys. J.}\ }\textbf {\bibinfo {volume} {966}},\ \bibinfo
  {pages} {186} (\bibinfo {year} {2024})},\ \Eprint
  {https://arxiv.org/abs/2402.04759} {arXiv:2402.04759 [astro-ph.HE]}
  \BibitemShut {NoStop}%
\end{thebibliography}%

\section*{End Matter}

In this appendix, we first provide details of the energy-calibration of $S_{38}$ and of $N_{68}$, and then explain the sources of energy-dependent systematic uncertainties in $N_{68}$ that enter into the budget $\sigma_C$ and justify the use of a non-zero $\delta C$ above 10\,EeV.

\begin{table}[h]
\caption{Coefficients of the second-degree polynomial in terms of $y_S=\log_{10}(S_{38}/(40~\textrm{VEM}))$ for the parameters $a_S$, $b_S$ and $c_S$.}
\label{tab:CICS1000}
\begin{ruledtabular}
\begin{tabular}{l c c c}
 & $~y_S^0~$ & $~y_S^1~$ & $~y_S^2~$ \\
\colrule
$~a_S~$ & $\phantom{+} 0.936$ & $\phantom{+} 0.005$ & $-0.400$  \\
$~b_S~$ & $-1.62\phantom{0}$ & $-0.51\phantom{0}$ & $-0.13\phantom{0}$  \\
$~c_S~$ & $-0.92\phantom{0}$ & $\phantom{+} 0.54\phantom{0} $ & $\phantom{+} 1.75\phantom{0} $  
\end{tabular}
\end{ruledtabular}
\end{table}

The energy calibration of $S(1000)$ is derived from a two-step process~\cite{PierreAuger:2020qqz}. Firstly,  $S(1000)$ is corrected for  attenuation effects with zenith angle $\theta$ by using the Constant Intensity Cut method~\cite{Hersil:1961zz}. For a given intensity threshold, the attenuation curve is fitted with a third-degree polynomial, $S(1000)=S_{38}(1+a_Sx_S+b_Sx_S^2+c_Sx_S^3)$, where  $x_S=\cos^2{\theta}-\cos^2{38^\circ}$ and $S_{38}$ is a zenith-independent shower-size estimator. The intensity-threshold dependence in the curves is accounted for by introducing an empirical dependence in terms of $y_S=\log_{10}(S_{38}/40~\textrm{VEM})$ in the coefficients $a_S$, $b_S$ and $c_S$ through a second-order polynomial in $y$. A VEM corresponds to the energy deposit of one vertical equivalent muon. The polynomial coefficients, updated with respect to those in~\cite{PierreAuger:2020qqz},  are given in Table~\ref{tab:CICS1000}. They relate to $S_{38}$ values ranging from 15~VEM to 120~VEM. Outside these bounds, the coefficients are set to their values at 15 and 120~VEM.  Secondly, the corrected shower-size estimator, $S_{38}$, is converted into energy $E$ using a power-law calibration relationship, $E=A_S(S_{38})^{B_S}$, determined with high-quality events detected simultaneously with the fluorescence detector. The energy scale is based on~\cite{Dawson:2019zva}, with recent improvements concerning aerosol attenuation~\cite{PierreAuger:2023nbk} and longitudinal-profile reconstruction~\cite{PierreAuger:2023att}. In this study, $A_S=(1.86\pm 0.03)\times 10^{-1}~$EeV and $B_S=1.021\pm 0.004$. 

\begin{table}[h]
\caption{Coefficients of the first-degree polynomial in terms of $y_N=\log_{10}N_{19})$ for the parameters $a_N$ and $b_N$.}
\label{tab:CICN19}
\begin{ruledtabular}
\begin{tabular}{l c c}
 & $~y_N^0~$ & $~y_N^1~$ \\ 
\colrule
$~a_N~$ & $-0.292$ & $\phantom{+} 0.468$  \\
$~b_N~$ & $-4.96\phantom{0}$ & $\phantom{+} 0.79\phantom{0}$ \\   
\end{tabular}
\end{ruledtabular}
\end{table}

The energy calibration of $N_{19}$ is carried out in the same manner. A constant-intensity-cut correction is applied to the (relative) muon-number estimator $N_{19}$ using a second-degree polynomial, $N_{19}=N_{68}(1+a_Nx_N+b_Nx_N^2)$, which is enough to guarantee the same intensity in bins of $\cos^2\theta$ between $60^\circ$ and $80^\circ$ (with $x_N=\cos^2\theta-\cos^268^\circ$). The intensity-threshold dependence is accounted for by introducing an empirical dependence in
terms of $y_N = \log_{10} (N_{19})$ in the coefficients $a_N$ and $b_N$. The polynomial coefficients are given in Table~\ref{tab:CICN19}. They relate to $N_{68}$ values ranging from $0.9$ to $4.5$. Outside these bounds, the
coefficients are set to their values at $0.9$ and $4.5$. The energy-calibration relationship is $E=A_N(N_{68})^{B_N}$, with $A_N=(5.29\pm 0.06)~$EeV and $B_N=1.046\pm 0.014$. Note that these values are refereed to as $A\pm\sigma_A$ and $B\pm\sigma_B$ in the main text. 

Since $N_{68}$ depends on the mass composition of the primary particles, its calibration relationship to the energy measured with the fluorescence technique accounts for the trend of the composition change with energy inherently as the underlying mass distribution is directly sampled by the fluorescence detectors. The use of a single power law for the relationship between $N_{68}$ and $E$ is then justified only for a single logarithmic evolution of the composition as a function of energy. This turns out to be the case, within the statistical uncertainties of the fluorescence dataset, above $\simeq 2~$EeV~\cite{PierreAuger:2014sui}. However, using deep-neural-network techniques applied to the surface-detector dataset, sharp changes of the elongation rate, too small to show up within the statistics available with the fluorescence technique, have recently been uncovered around $\simeq 6$\,EeV, $\simeq 10$\,EeV and $\simeq 30$\,EeV~\cite{PierreAuger:2024flk,PierreAuger:2024nzw}. The impact of these elongation rate changes on the energy calibration of $N_{68}$ can be estimated as follows. Within the Heitler-Matthews superposition model of air shower~\cite{Matthews:2005sd}, the number of muons $N_\upmu$ increases with the cosmic-ray mass number $A$ as $N_\upmu\propto A^{1-\beta}$. Assuming that the slope of $N_{68}$ with $A$ follows that of $N_\upmu$, the change of slope of $N_{68}$, $\Delta m_{N_{68}}$, can be related to that of the shower maximum slant depth, $\Delta m_{X_\mathrm{max}}$ through
\begin{equation}
    \Delta m_{N_{68}}=-\frac{1-\beta}{D_0}\Delta m_{X_\mathrm{max}},
\end{equation}
with $D_0\simeq 56.1~$g~cm$^{-2}$ the  elongate rate of protons expected from EPOS-LHC~\cite{Pierog:2013ria}. Using $\beta\simeq 0.927$ as a benchmark~\cite{Cazon:2018gww}, and considering to first order a single change of elongation rate around 10\,EeV to describe the observed series of kinks from below 6\,EeV to above 30\,EeV, we get an amplitude for the expected non-linearity of $N_{68}$ of $\simeq 2.5\%$ per decade. We note that we neglect here the correlated non-linearity effects in the energy calibration of $S_{38}$ as they are smaller.
 
Another source of logarithmic non-linearities between $N_{68}$ and $E$ stems from the use of the constant-intensity-cut correction to make uniform the distribution of events in terms of $\sin^2\theta$. The correction is necessary to compensate for imperfections of the $N_{19}$ energy estimator, which accounts for muon contributions to the signals and requires model-dependent corrections for electromagnetic ones, not negligible between $\theta=60^\circ$ and $\theta\simeq 70^\circ$. However, smaller, yet genuine, deviations of the $\sin^2\theta$ distribution from uniformity, $\Delta$, are expected in presence of dipolar anisotropies. They can be estimated as~\cite{PierreAuger:2012xju}
\begin{equation}
    \Delta=\frac{N_\mathrm{dip}}{N_\mathrm{iso}}d_z\sin\lambda\cos\theta,
\end{equation}
with $N_\mathrm{iso}$ ($N_\mathrm{dip}$) the expected number of events in the covered region of the sky for an isotropic (a dipolar) angular distribution. While $\Delta$ is well within 1\% for $\theta\leq60^\circ$ for the dipole components $d_z$ of interest (measured with large uncertainties, though), it increases up to, on average, $\simeq 3$\% for $\theta\geq60^\circ$. We estimated that the uniformity of the $\sin^2\theta$ distribution forced by the constant intensity correction for $N_{68}<4.5$ ($E\simeq 20~$EeV) can then lead to energy distortions of $\simeq 1.5\%$ between 4 and $\simeq 20$\,EeV, while higher energies would not be impacted as the energy dependence of the polynomial coefficients relating $N_{19}$ to $N_{68}$ is frozen. This effectively acts as a source of non-linearity that occurs around $\simeq 20$\,EeV. 

Overall, we add in quadrature both sources of non-linearities and end up with $\sigma_C=3\times 10^{-2}$, and choose, as a first-order approximation, the transition energy at 10\,EeV. We have checked that the best-fit value for $\delta C$ is mildly impacted when increasing $\sigma_C$ to $5\times 10^{-2}$ or increasing the transition energy.

\section*{Supplemental material}

We provide in this supplemental material a few summary plots and data related to the analysis presented in the article.  

\begin{figure}[b]
\centering
\includegraphics[width=0.5\textwidth]{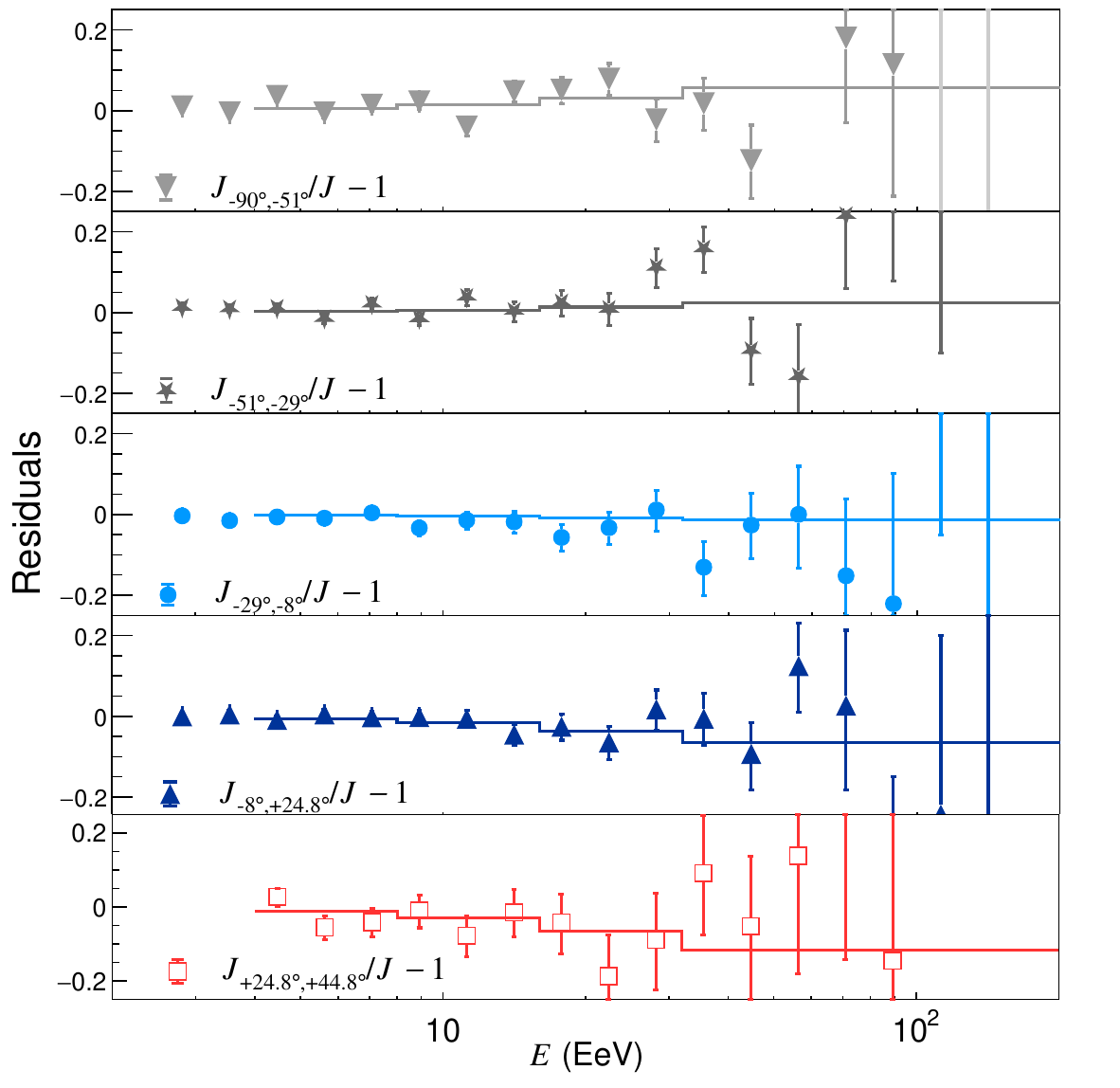}  
\caption{Energy-spectra residuals zoomed within $\pm25\%$ in the five declination bands.}
\label{fig:residuals_bis}
\end{figure}

The response function for events between 0 and $60^\circ$ is described in detail in our previous paper~\cite{PierreAuger:2020qqz}.  We have applied the same data-driven techniques to determine that for events  between $60^\circ$ and $80^\circ$. The resolution function is parameterized as a Gaussian function with $\sigma_N$ parameter evolving with energy as
\begin{equation}
    \frac{\sigma_N(E)}{E}=0.046+\frac{0.34}{\sqrt{E}},
\end{equation}
with $E$ in EeV. No significant bias is found in the sub-threshold energy bins of interest, down to 2.5\,EeV. The efficiency is described as (still with $E$ in EeV)
\begin{equation}
    \epsilon(E)=0.5\left(1+\mathrm{erf}\left(\frac{\log_{10}E}{0.34}\right)\right).
\end{equation}

In Fig.~\ref{fig:residuals_bis}, we present an alternative version of the bottom panel of Fig.~\ref{fig:residuals}. Here, the energy-spectra residuals in the five declination bands are zoomed within $\pm25\%$ while no artificial shifts are applied for visualization purpose;  some data points at high energies therefore lie outside the windows. The reference lines are the best-fit function for the spectrum combined over $[-84.8^\circ,+24.8^\circ]$ accounting for dipole anisotropies.

The correlation matrix relative to the covariance matrices used to search for differences in the spectral features is given in Table~\ref{tab:covmat}. It provides, together with the corresponding uncertainties in the parameters listed in Table~\ref{tab:features}, the coefficients needed to get $\mathbf{\Sigma}_{\mathrm{ref}}$ as well as $\mathbf{\Sigma}_{\Delta\delta}$. 

The combined spectrum data points with their statistical and systematic uncertainties are collected in Table~\ref{tab:J} together with the observed and corrected number of events, while the correlation matrix of the spectrum data points that accounts for systematic uncertainties is given in Table~\ref{tab:rhoJ}. The strong bin-to-bin correlations observed stem from systematic uncertainties dominated by those in the energy scale: for a single power-law spectrum in $E^{-\gamma}$, a $\Delta E/E$ change in energy scale translates indeed as a change $\Delta J(E)/J(E)=(\gamma-1) \Delta E/E$. The correlation matrix together with the uncertainties collected in Table~\ref{tab:J} allow for a direct model/data comparison through a generalized $\chi^2$ that accounts for all sources of uncertainties. It can also be used to propagate uncertainties in any quantity derived from the cosmic-ray spectrum, such as cosmogenic gamma-ray~\cite{Berat:2022iea} or neutrino fluxes~\cite{Berat:2024rvf}.

Finally, the spectra obtained in each declination band are collected in Tables~\ref{tab:J1}-to-\ref{tab:J5}.

\begin{table}[!t]
\caption{Correlation matrix relative to $\mathbf{\Sigma}_{\mathrm{ref}}$ expressed in a basis $(J_0,\gamma_1,E_{12},\gamma_2,E_{23},\gamma_3,E_{34},\gamma_4)$.}
\label{tab:covmat}
\begin{ruledtabular}
\begin{tabular}{c c c c c c c c}
\phantom{+}1.000 & $-0.535$ & $+0.613$ & $-0.396$  & $-0.162$ & $+0.218$ & $-0.182$ & $-0.317$ \\ 
 & \phantom{+}1.000  & $-0.782$ & $+0.494$ & $+0.343$ & $+0.111$ & $+0.561$ & $+0.512$ \\
 & & \phantom{+}1.000 & $-0.872$ & $-0.609$ & $-0.186$ & $-0.329$ & $-0.303$ \\
 & & & \phantom{+}1.000  & $+0.827$ & $+0.313 $ & $+0.104$ &$+0.052$ \\
 & & & & \phantom{+}1.000 & $+0.594$ & $+0.007$ & $-0.052$\\
 & & & & & \phantom{+}1.000 & $+0.133$ & $-0.186$\\
 & & & & & & \phantom{+}1.000  & $+0.431$\\
 & & & & & & & \phantom{+}1.000\\
&\\[-1.0em]
 
\end{tabular}
\end{ruledtabular}
\end{table}

\begin{table*}[h]
\caption{Combined spectrum data.}
\label{tab:J}
\begin{ruledtabular}
\begin{tabular}{l c c c}
 $\log_{10} (E/\mathrm{EeV})$ & $J(E) \pm \sigma_\mathrm{stat}(E) \pm \sigma_\mathrm{syst}(E)$ (EeV$^{-1}$~km$^{-2}$~sr$^{-1}$~yr$^{-1}$) & $n$ & $n_\mathrm{corr}$\\
\colrule
&\\[-1.0em]
0.45 & $\left(1.8614^{~+0.0065 ~~+0.7}_{~-0.0065 ~-0.6}\right) \times 10^{0}$ & 107232 & 98178.2 \\
&&&\\[-1.0em]
0.55 & $\left(8.808^{~+0.040 ~~+3.2}_{~-0.040 ~-2.5}\right) \times 10^{-1}$ & 61873 & 58483.8 \\
&&&\\[-1.0em]
0.65 & $\left(4.183^{~+0.022 ~~+1.4}_{~-0.022 ~-1.1}\right) \times 10^{-1}$ & 48929 & 45207.8  \\
&&&\\[-1.0em]
0.75 & $\left(2.107^{~+0.014 ~~+0.6}_{~-0.014 ~-0.4}\right) \times 10^{-1}$ & 30525 & 28678.2  \\
&&&\\[-1.0em]
0.85 & $\left(1.186^{~+0.009 ~~+0.3}_{~-0.009~-0.3}\right) \times 10^{-1}$ & 20984 & 20314.9  \\
&&&\\[-1.0em]
0.95 & $\left(6.590^{~+0.059 ~~+1.5}_{~-0.059 ~-1.4}\right) \times 10^{-2}$ & 14590 & 14213.9  \\
&&&\\[-1.0em]
1.05 & $\left(3.668^{~+0.039 ~~+0.9}_{~-0.039 ~-0.8}\right) \times 10^{-2}$ & 10159 & 9957.9 \\
&&&\\[-1.0em]
1.15 & $\left(2.001^{~+0.024 ~~+0.5}_{~-0.026 ~-0.6}\right) \times 10^{-2}$ & 6984 & 6858.1  \\
&&&\\[-1.0em]
1.25 & $\left(9.982^{~+0.152 ~~+3.0}_{~-0.165 ~-2.8}\right) \times 10^{-3}$ & 4396 & 4296.4  \\
&&&\\[-1.0em]
1.35 & $\left(5.027^{~+0.097 ~~+1.5}_{~-0.101 ~-1.3}\right) \times 10^{-3}$ & 2775 & 2724.0  \\
&&&\\[-1.0em]
1.45 & $\left(2.600^{~+0.062 ~~+0.7}_{~-0.067~-0.7}\right) \times 10^{-3}$ & 1800 & 1773.7  \\
&&&\\[-1.0em]
1.55 & $\left(1.294^{~+0.039 ~~+0.4}_{~-0.043 ~-0.3}\right) \times 10^{-3}$ & 1124 & 1112.3  \\
&&&\\[-1.0em]
1.65 & $\left(5.88^{~+0.23 ~~+2.7}_{~-0.26 ~-2.4}\right) \times 10^{-4}$ & 634 & 635.3 \\
&&&\\[-1.0em]
1.75 & $\left(2.27^{~+0.13 ~~+1.2}_{~-0.14 ~-1.1}\right) \times 10^{-4}$ & 318 & 309.3   \\
&&&\\[-1.0em]
1.85 & $\left(6.98^{~+0.67 ~~+5.3}_{~-0.69 ~-3.4}\right) \times 10^{-5}$ & 126 & 119.5   \\
&&&\\[-1.0em]
1.95 & $\left(1.89^{~+0.29 ~~+1.9}_{~-0.32 ~-0.6}\right) \times 10^{-5}$ & 43 & 40.8  \\
&&&\\[-1.0em]
2.05 & $\left(5.3^{~+1.4 ~~+6.0}_{~-1.6 ~-2.8}\right) \times 10^{-6}$ & 15 & 14.2  \\
&&&\\[-1.0em]
2.15 & $\left(2.5^{~+0.7 ~~+0.1}_{~-1.1 ~-1.9}\right) \times 10^{-6}$ & 9 & 8.5 \\
&&&\\[-1.0em]
2.25 & $< 5.4 \times 10^{-7}$ & 0 & 0 \\
&&&\\[-1.0em]
2.35 & $<  4.3\times 10^{-7}$ & 0 & 0 \\
\end{tabular}
\end{ruledtabular}
\end{table*}

\begin{table*}[h]
\caption{Correlation matrix of the spectrum accounting for systematic uncertainties.}
\label{tab:rhoJ}
\begin{ruledtabular}
\begin{tabular}{l c c c c c c c c c c c c c c c c c c}
 \diagbox{$i$}{$j$} & 0.45 & 0.55 & 0.65 & 0.75 & 0.85 & 0.95 & 1.05 & 1.15 & 1.25 & 1.35 & 1.45 & 1.55 & 1.65 & 1.75 & 1.85 & 1.95 & 2.05 & 2.15 \\
\colrule
&\\[-1.0em]
0.45 & 1 & 0.984 & 0.983 & 0.980 & 0.985 & 0.991 & 0.986 & 0.976 & 0.971 & 0.970 & 0.980 & 0.987 & 0.990 & 0.983 & 0.974 & 0.946 & 0.957 & 0.788 \\ 
&&&\\[-1.0em]
0.55 &  & 1 & 0.999 &  0.999 & 0.993 & 0.993 & 0.986 & 0.977 & 0.978 & 0.977 & 0.981 & 0.986 & 0.990 & 0.997 & 0.994  & 0.973  & 0.963 & 0.848 \\ 
&&&\\[-1.0em]
0.65 & & & 1 & 0.999 & 0.994 & 0.993 & 0.984 & 0.975  & 0.977 & 0.975 & 0.983 & 0.987 & 0.991 &  0.997 & 0.994 & 0.972 & 0.962 & 0.853 \\ 
&&&\\[-1.0em]
0.75 & & &  & 1 & 0.993 & 0.991 & 0.982 & 0.970 &  0.971 & 0.970 &  0.978 & 0.987 & 0.989 & 0.997 & 0.994 & 0.974 & 0.960 &  0.851 \\ 
&&&\\[-1.0em]
0.85 &  & & & & 1 & 0.997 & 0.986 & 0.977 & 0.967  & 0.965 & 0.986 & 0.996 & 0.997 & 0.992 & 0.982 & 0.949 &  0.955 & 0.859 \\ 
&&&\\[-1.0em]
0.95 &  & & & & & 1 & 0.992 & 0.985 & 0.977 & 0.974 & 0.986&  0.994 &  0.994 & 0.990 &  0.981 &  0.953 & 0.951 & 0.842 \\ 
&&&\\[-1.0em]
1.05 & & & & & & & 1 & 0.994 & 0.991 & 0.989 & 0.980 &  0.985 & 0.982 & 0.976 & 0.971 & 0.945 & 0.949 & 0.805 \\ 
&&&\\[-1.0em]
1.15 & & & & & & & & 1 & 0.993 & 0.992 & 0.981 & 0.973 & 0.974 & 0.964 & 0.958 & 0.924 & 0.941 & 0.813 \\ 
&&&\\[-1.0em]
1.25 & & & & & & & & & 1 & 0.998 & 0.976 & 0.964 & 0.963 & 0.964 & 0.965&  0.943 & 0.937 & 0.803 \\ 
&&&\\[-1.0em]
1.35 & & & & & & & & & & 1 & 0.973 & 0.961 & 0.963 & 0.962 & 0.962 & 0.942 & 0.943 & 0.787 \\ 
&&&\\[-1.0em]
1.45 & & & & & & & & & & & 1 &  0.986 & 0.984 & 0.974 & 0.963 & 0.922 & 0.941 & 0.850 \\ 
&&&\\[-1.0em]
1.55 &  & & & & & & & & & & & 1 & 0.993 & 0.983 & 0.974 & 0.937 & 0.944 & 0.842 \\ 
&&&\\[-1.0em]
1.65 & & & & & & & & & & & & & 1 & 0.989 & 0.978 & 0.946 & 0.962 & 0.832 \\ 
&&&\\[-1.0em]
1.75 & & & & & & & & & & & & & & 1 & 0.994 & 0.974 & 0.959 & 0.852 \\ 
&&&\\[-1.0em]
1.85 & & & & & & & & & & & & & & & 1 & 0.981 &  0.949 & 0.849 \\ 
&&&\\[-1.0em]
1.95 &  & & & & & & & & & & & & & & & 1 & 0.923 & 0.785 \\ 
&&&\\[-1.0em]
2.05 & & & & & & & & & & & & & & & &  & 1 & 0.756 \\ 
&&&\\[-1.0em]
2.15 & & & & & & & & & & & & & & & & & & 1 \\ 
\end{tabular}
\end{ruledtabular}
\end{table*}

\begin{table*}[h]
\caption{Spectrum data over the declination band $[-90^\circ,-51^\circ]$.}
\label{tab:J1}
\begin{ruledtabular}
\begin{tabular}{l c c c}
 $\log_{10} (E/\mathrm{EeV})$ & $J(E) \pm \sigma_\mathrm{stat}(E) $ (EeV$^{-1}$~km$^{-2}$~sr$^{-1}$~yr$^{-1}$) & $n$ & $n_\mathrm{corr}$\\
\colrule
&\\[-1.0em]
0.45 & $\left(1.875_{~-0.013}^{~+0.013}\right) \times 10^{0}$ & 24889 & 22728.5 \\
&&&\\[-1.0em]
0.55 & $\left(8.797_{~-0.081}^{~+0.082}\right) \times 10^{-1}$ & 14198 & 13424.9 \\
&&&\\[-1.0em]
0.65 & $\left(4.294_{~-0.045}^{~+0.041}\right) \times 10^{-1}$ & 11557 & 10693.2 \\
&&&\\[-1.0em]
0.75 & $\left(2.114_{~-0.028}^{~+0.026}\right) \times 10^{-1}$ & 7045 & 6628.6 \\
&&&\\[-1.0em]
0.85 & $\left(1.194_{~-0.019}^{~+0.017}\right) \times 10^{-1}$ & 4873 & 4714.3 \\
&&&\\[-1.0em]
0.95 & $\left(6.793_{~-0.126}^{~+0.116}\right) \times 10^{-2}$ & 3470 & 3375.1 \\
&&&\\[-1.0em]
1.05 & $\left(3.549_{~-0.082}^{~+0.075}\right) \times 10^{-2}$ & 2278 & 2220.4 \\
&&&\\[-1.0em]
1.15 & $\left(2.104_{~-0.057}^{~+0.052}\right) \times 10^{-2}$ & 1685 & 1657.4 \\
&&&\\[-1.0em]
1.25 & $\left(1.057_{~-0.035}^{~+0.032}\right) \times 10^{-2}$ & 1057 & 1048.1\\
&&&\\[-1.0em]
1.35 & $\left(5.49_{~-0.22}^{~+0.21}\right) \times 10^{-3}$ & 699 & 684.7 \\
&&&\\[-1.0em]
1.45 & $\left(2.48_{~-0.14}^{~+0.13}\right) \times 10^{-3}$ & 397 & 389.4 \\
&&&\\[-1.0em]
1.55 & $\left(1.30_{~-0.09}^{~+0.08}\right) \times 10^{-3}$ & 263 & 258.0 \\
&&&\\[-1.0em]
1.65 & $\left(5.54_{~-0.52}^{~+0.49}\right) \times 10^{-4}$ & 141 & 137.9 \\
&&&\\[-1.0em]
1.75 & $\left(2.78_{~-0.31}^{~+0.30}\right) \times 10^{-4}$ & 85 & 86.8 \\
&&&\\[-1.0em]
1.85 & $\left(7.7_{~-1.6}^{~+1.4}\right) \times 10^{-5}$ & 31 & 30.3 \\
&&&\\[-1.0em]
1.95 & $\left(2.1_{~-0.7}^{~+0.7}\right) \times 10^{-5}$ & 12 & 10.4 \\
&&&\\[-1.0em]
2.05 & $\left(1.4_{~-1.2}^{~+1.8}\right) \times 10^{-6}$ & 1 & 0.9 \\
&&&\\[-1.0em]
2.15 & $\left(1.1_{~-0.9}^{~+1.4}\right) \times 10^{-6}$ & 1 & 0.9 \\
&&&\\[-1.0em]
2.25 & $<2.0 \times 10^{-6}$ & 0 & 0 \\
&&&\\[-1.0em]
2.35 & $<1.6 \times 10^{-6}$ & 0 & 0 \\
\end{tabular}
\end{ruledtabular}
\end{table*}

\begin{table*}[h]
\caption{Spectrum data over the declination band $[-51^\circ,-29^\circ]$.}
\label{tab:J2}
\begin{ruledtabular}
\begin{tabular}{l c c c}
 $\log_{10} (E/\mathrm{EeV})$ & $J(E) \pm \sigma_\mathrm{stat}(E) $ (EeV$^{-1}$~km$^{-2}$~sr$^{-1}$~yr$^{-1}$) & $n$ & $n_\mathrm{corr}$\\
\colrule
&\\[-1.0em]
0.45 & $\left(1.881_{~-0.012}^{~+0.012}\right) \times 10^{0}$ & 29491 & 27226.2 \\
&&&\\[-1.0em]
0.55 & $\left(8.921_{~-0.076}^{~+0.076}\right) \times 10^{-1}$ & 17193 & 16258.8 \\
&&&\\[-1.0em]
0.65 & $\left(4.192_{~-0.043}^{~+0.039}\right) \times 10^{-1}$ & 12052 & 11194.9 \\
&&&\\[-1.0em]
0.75 & $\left(2.095_{~-0.027}^{~+0.025}\right) \times 10^{-1}$ & 7509 & 7044.0 \\
&&&\\[-1.0em]
0.85 & $\left(1.202_{~-0.019}^{~+0.017}\right) \times 10^{-1}$ & 5247 & 5088.8 \\
&&&\\[-1.0em]
0.95 & $\left(6.544_{~-0.119}^{~+0.110}\right) \times 10^{-2}$ & 3565 & 3487.2 \\
&&&\\[-1.0em]
1.05 & $\left(3.836_{~-0.082}^{~+0.075}\right) \times 10^{-2}$ & 2616 & 2573.7 \\
&&&\\[-1.0em]
1.15 & $\left(2.015_{~-0.052}^{~+0.048}\right) \times 10^{-2}$ & 1743 & 1702.3 \\
&&&\\[-1.0em]
1.25 & $\left(1.030_{~-0.033}^{~+0.031}\right) \times 10^{-2}$ & 1118 & 1095.3 \\
&&&\\[-1.0em]
1.35 & $\left(5.13_{~-0.21}^{~+0.20}\right) \times 10^{-3}$ & 697 & 686.8 \\
&&&\\[-1.0em]
1.45 & $\left(2.82_{~-0.14}^{~+0.13}\right) \times 10^{-3}$ & 479 & 474.8 \\
&&&\\[-1.0em]
1.55 & $\left(1.48_{~-0.09}^{~+0.08}\right) \times 10^{-3}$ & 313 & 314.9 \\
&&&\\[-1.0em]
1.65 & $\left(5.72_{~-0.48}^{~+0.46}\right) \times 10^{-4}$ & 157 & 152.9 \\
&&&\\[-1.0em]
1.75 & $\left(1.84_{~-0.24}^{~+0.24}\right) \times 10^{-4}$ & 65 & 62.0 \\
&&&\\[-1.0em]
1.85 & $\left(8.1_{~-1.5}^{~+1.4}\right) \times 10^{-5}$ & 35 & 34.2 \\
&&&\\[-1.0em]
1.95 & $\left(2.7_{~-0.7}^{~+0.8}\right) \times 10^{-5}$ & 15 & 14.7 \\
&&&\\[-1.0em]
2.05 & $\left(7.3_{~-2.8}^{~+3.2}\right) \times 10^{-6}$ & 5 & 4.9 \\
&&&\\[-1.0em]
2.15 & $\left(5.8_{~-2.3}^{~+2.5}\right) \times 10^{-6}$ & 5 & 4.9 \\
&&&\\[-1.0em]
2.25 & $<2.1 \times 10^{-6}$ & 0 & 0 \\
&&&\\[-1.0em]
2.35 & $<1.7 \times 10^{-6}$ & 0 & 0 \\
\end{tabular}
\end{ruledtabular}
\end{table*}

\begin{table*}[h]
\caption{Spectrum data over the declination band $[-29^\circ,-8^\circ]$.}
\label{tab:J3}
\begin{ruledtabular}
\begin{tabular}{l c c c}
 $\log_{10} (E/\mathrm{EeV})$ & $J(E) \pm \sigma_\mathrm{stat}(E) $ (EeV$^{-1}$~km$^{-2}$~sr$^{-1}$~yr$^{-1}$) & $n$ & $n_\mathrm{corr}$\\
\colrule
&\\[-1.0em]
0.45 & $\left(1.850_{~-0.012}^{~+0.012}\right) \times 10^{-0}$ & 28269 & 25961.0 \\
&&&\\[-1.0em]
0.55 & $\left(8.703_{~-0.076}^{~+0.076}\right) \times 10^{-1}$ & 16303 & 15376.7 \\
&&&\\[-1.0em]
0.65 & $\left(4.119_{~-0.043}^{~+0.039}\right) \times 10^{-1}$ & 11529 & 10677.9 \\
&&&\\[-1.0em]
0.75 & $\left(2.105_{~-0.027}^{~+0.025}\right) \times 10^{-1}$ & 7258 & 6870.6 \\
&&&\\[-1.0em]
0.85 & $\left(1.182_{~-0.018}^{~+0.017}\right) \times 10^{-1}$ & 5003 & 4854.1 \\
&&&\\[-1.0em]
0.95 & $\left(6.420_{~-0.119}^{~+0.111}\right) \times 10^{-2}$ & 3408 & 3319.7 \\
&&&\\[-1.0em]
1.05 & $\left(3.642_{~-0.080}^{~+0.075}\right) \times 10^{-2}$ & 2420 & 2371.2 \\
&&&\\[-1.0em]
1.15 & $\left(1.972_{~-0.054}^{~+0.049}\right) \times 10^{-2}$ & 1646 & 1615.9 \\
&&&\\[-1.0em]
1.25 & $\left(9.482_{~-0.323}^{~+0.296}\right) \times 10^{-3}$ & 1001 & 978.3 \\
&&&\\[-1.0em]
1.35 & $\left(4.92_{~-0.21}^{~+0.19}\right) \times 10^{-3}$ & 651 & 639.1 \\
&&&\\[-1.0em]
1.45 & $\left(2.56_{~-0.14}^{~+0.13}\right) \times 10^{-3}$ & 425 & 419.0 \\
&&&\\[-1.0em]
1.55 & $\left(1.11_{~-0.08}^{~+0.07}\right) \times 10^{-3}$ & 232 & 229.3 \\
&&&\\[-1.0em]
1.65 & $\left(6.15_{~-0.52}^{~+0.49}\right) \times 10^{-4}$ & 159 & 159.4 \\
&&&\\[-1.0em]
1.75 & $\left(2.19_{~-0.29}^{~+0.26}\right) \times 10^{-4}$ & 73 & 71.5 \\
&&&\\[-1.0em]
1.85 & $\left(5.5_{~-1.1}^{~+1.0}\right) \times 10^{-5}$ & 24 & 22.7 \\
&&&\\[-1.0em]
1.95 & $\left(1.5_{~-0.5}^{~+0.5}\right) \times 10^{-5}$ & 8 & 7.5 \\
&&&\\[-1.0em]
2.05 & $\left(8.7_{~-4.1}^{~+3.6}\right) \times 10^{-6}$ & 6 & 5.7 \\
&&&\\[-1.0em]
2.15 & $\left(1.2_{~-1.0}^{~+1.3}\right) \times 10^{-6}$ & 1 & 0.9 \\
&&&\\[-1.0em]
2.25 & $<2.2 \times 10^{-6}$ & 0 & 0 \\
&&&\\[-1.0em]
2.35 & $<1.7 \times 10^{-6}$ & 0 & 0 \\
\end{tabular}
\end{ruledtabular}
\end{table*}

\begin{table*}[h]
\caption{Spectrum data over the declination band $[-8^\circ,+24.8^\circ]$.}
\label{tab:J4}
\begin{ruledtabular}
\begin{tabular}{l c c c}
 $\log_{10} (E/\mathrm{EeV})$ & $J(E) \pm \sigma_\mathrm{stat}(E) $ (EeV$^{-1}$~km$^{-2}$~sr$^{-1}$~yr$^{-1}$) & $n$ & $n_\mathrm{corr}$\\
\colrule
&\\[-1.0em]
0.45 & $\left(1.857_{~-0.013}^{~+0.013}\right) \times 10^{0}$ & 24477 & 22123.1 \\
&&&\\[-1.0em]
0.55 & $\left(8.880_{~-0.083}^{~+0.083}\right) \times 10^{-1}$ & 14109 & 13320.9 \\
&&&\\[-1.0em]
0.65 & $\left(4.109_{~-0.043}^{~+0.039}\right) \times 10^{-1}$ & 11712 & 10776.8 \\
&&&\\[-1.0em]
0.75 & $\left(2.136_{~-0.028}^{~+0.025}\right) \times 10^{-1}$ & 7523 & 7055.6 \\
&&&\\[-1.0em]
0.85 & $\left(1.175_{~-0.018}^{~+0.017}\right) \times 10^{-1}$ & 5048 & 4885.2 \\
&&&\\[-1.0em]
0.95 & $\left(6.632_{~-0.122}^{~+0.111}\right) \times 10^{-2}$ & 3560 & 3471.6 \\
&&&\\[-1.0em]
1.05 & $\left(3.673_{~-0.080}^{~+0.074}\right) \times 10^{-2}$ & 2464 & 2420.3 \\
&&&\\[-1.0em]
1.15 & $\left(1.918_{~-0.053}^{~+0.048}\right) \times 10^{-2}$ & 1634 & 1591.4 \\
&&&\\[-1.0em]
1.25 & $\left(9.796_{~-0.327}^{~+0.305}\right) \times 10^{-3}$ & 1049 & 1023.0 \\
&&&\\[-1.0em]
1.35 & $\left(4.76_{~-0.21}^{~+0.19}\right) \times 10^{-3}$ & 638 & 625.9 \\
&&&\\[-1.0em]
1.45 & $\left(2.57_{~-0.14}^{~+0.12}\right) \times 10^{-3}$ & 433 & 426.6 \\
&&&\\[-1.0em]
1.55 & $\left(1.27_{~-0.08}^{~+0.07}\right) \times 10^{-3}$ & 269 & 265.4 \\
&&&\\[-1.0em]
1.65 & $\left(5.72_{~-0.51}^{~+0.45}\right) \times 10^{-4}$ & 150 & 150.2 \\
&&&\\[-1.0em]
1.75 & $\left(2.46_{~-0.28}^{~+0.26}\right) \times 10^{-4}$ & 82 & 81.4 \\
&&&\\[-1.0em]
1.85 & $\left(6.7_{~-1.4}^{~+1.3}\right) \times 10^{-5}$ & 30 & 27.8 \\
&&&\\[-1.0em]
1.95 & $\left(1.0_{~-0.5}^{~+0.5}\right) \times 10^{-5}$ & 6 & 5.5 \\
&&&\\[-1.0em]
2.05 & $\left(4.2_{~-2.7}^{~+1.9}\right) \times 10^{-6}$ & 3 & 2.8 \\
&&&\\[-1.0em]
2.15 & $\left(2.2_{~-2.1}^{~+1.4}\right) \times 10^{-6}$ & 2 & 1.8 \\
&&&\\[-1.0em]
2.25 & $<2.0 \times 10^{-6}$ & 0 & 0 \\
&&&\\[-1.0em]
2.35 & $<1.6 \times 10^{-6}$ & 0 & 0 \\
\end{tabular}
\end{ruledtabular}
\end{table*}

\begin{table*}[h]
\caption{Spectrum data over the declination band $[+24.8^\circ,+44.8^\circ]$.}
\label{tab:J5}
\begin{ruledtabular}
\begin{tabular}{l c c c}
 $\log_{10} (E/\mathrm{EeV})$ & $J(E) \pm \sigma_\mathrm{stat}(E) $ (EeV$^{-1}$~km$^{-2}$~sr$^{-1}$~yr$^{-1}$) & $n$ & $n_\mathrm{corr}$\\
\colrule
&\\[-1.0em]
0.65 & $\left(4.259_{~-0.109}^{~+0.100}\right) \times 10^{-1}$ & 2037 & 1793.8 \\
&&&\\[-1.0em]
0.75 & $\left(2.008_{~-0.066}^{~+0.061}\right) \times 10^{-1}$ & 1172 & 1064.9 \\
&&&\\[-1.0em]
0.85 & $\left(1.13_{~-0.04}^{~+0.04}\right) \times 10^{-1}$ & 793 & 753.6 \\
&&&\\[-1.0em]
0.95 & $\left(6.58_{~-0.31}^{~+0.28}\right) \times 10^{-2}$ & 574 & 552.7 \\
&&&\\[-1.0em]
1.05 & $\left(3.41_{~-0.19}^{~+0.18}\right) \times 10^{-2}$ & 371 & 360.6 \\
&&&\\[-1.0em]
1.15 & $\left(1.98_{~-0.13}^{~+0.13}\right) \times 10^{-2}$ & 270 & 263.7 \\
&&&\\[-1.0em]
1.25 & $\left(9.63_{~-0.81}^{~+0.74}\right) \times 10^{-3}$ & 167 & 161.6 \\
&&&\\[-1.0em]
1.35 & $\left(4.13_{~-0.49}^{~+0.46}\right) \times 10^{-3}$ & 90 & 87.3 \\
&&&\\[-1.0em]
1.45 & $\left(2.31_{~-0.31}^{~+0.29}\right) \times 10^{-3}$ & 63 & 61.4 \\
&&&\\[-1.0em]
1.55 & $\left(1.4_{~-0.2}^{~+0.2}\right) \times 10^{-3}$ & 48 & 46.8 \\
&&&\\[-1.0em]
1.65 & $\left(6.0_{~-1.2}^{~+1.1}\right) \times 10^{-4}$ & 26 & 25.2 \\
&&&\\[-1.0em]
1.75 & $\left(2.5_{~-0.8}^{~+0.75}\right) \times 10^{-4}$ & 13 & 13.2 \\
&&&\\[-1.0em]
1.85 & $\left(9.1_{~-4.6}^{~+4.3}\right) \times 10^{-5}$ & 6 & 6.1 \\
&&&\\[-1.0em]
1.95 & $\left(1.6_{~-1.4}^{~+0.8}\right) \times 10^{-5}$ & 2 & 1.3 \\
&&&\\[-1.0em]
2.05 & $<1.2  \times 10^{-5}$ & 0 & 0 \\
&&&\\[-1.0em]
2.15 & $<9.8  \times 10^{-6}$ & 0 & 0 \\
&&&\\[-1.0em]
2.25 & $<8.0 \times 10^{-6}$ & 0 & 0 \\
&&&\\[-1.0em]
2.35 & $<6.5 \times 10^{-6}$ & 0 & 0 \\
\end{tabular}
\end{ruledtabular}
\end{table*}

\end{document}